\documentclass[a4paper, halfparskip]{scrartcl}

\RequirePackage[T1]{fontenc}
\RequirePackage{lmodern}
\usepackage[utf8]{inputenc}
\usepackage{textcomp,amssymb}
\usepackage[babel]{microtype}
\usepackage{graphicx}
\usepackage{epstopdf}
\usepackage{pdfpages}
\usepackage{placeins}
\usepackage{siunitx}
\usepackage{todonotes}
\usepackage{csquotes}
\usepackage{verbatim}
\usepackage{multicol}
\usepackage{multirow}
\MakeOuterQuote{"}

\usepackage{cite}
\usepackage{bibgerm}

\usepackage{ucs}

\usepackage{amsmath, amsthm, amssymb}
\usepackage{amsfonts}
\usepackage{subfig}
\usepackage{nomencl}
\usepackage{bibgerm}
\usepackage{bm}

\usepackage{calrsfs}
\DeclareMathAlphabet{\pazocal}{OMS}{zplm}{m}{n}

\usepackage{array} 
\usepackage{cancel}

\usepackage{exscale}
\usepackage{longtable} 
\usepackage{booktabs}
\usepackage{enumitem}
\usepackage{comment}
\usepackage{graphicx}

\usepackage{tikz}

\usepackage{stackengine}
\newcommand\xrowht[2][0]{\addstackgap[.5\dimexpr#2\relax]{\vphantom{#1}}}

\usepackage[toc,page]{appendix}

\usepackage{ulem}

\begin{document}
\selectlanguage{english}
\begin{titlepage}
   \begin{center}
       \vspace*{1cm}

       \textbf{Adjoint-based Shape Optimization for the Minimization of Flow-induced Hemolysis in Biomedical Applications}
       \vspace{0.5cm}
       \\Georgios Bletsos\footnote{george.bletsos@tuhh.de}, Niklas Kühl and Thomas Rung
        \vspace{0.5cm}
       \\Hamburg University of Technology, Institute for Fluid Dynamics and Ship Theory, Am Schwarzenberg-Campus 4, D-21075 Hamburg, Germany
       \vspace{0.3cm}
       \\January, 2021
       
       \vspace{1.5cm}

        \textbf{Abstract}
        \end{center}
        \par 
        This paper reports on the derivation and implementation of a shape optimization procedure for the minimization of hemolysis induction in biomedical devices.
        Hemolysis is a blood damaging phenomenon that may occur in 
        mechanical blood-processing applications where large velocity gradients are found. An increased level of damaged blood can lead to deterioration of the immune system and quality of life. 
        It is, thus, important to minimize flow-induced hemolysis by improving the design of next-generation biomedical machinery.
        Emphasis is given to the formulation of a continuous adjoint complement to a power-law hemolysis prediction model dedicated to efficiently identifying the shape sensitivity to hemolysis.
        The computational approach is verified against the analytical solutions of a benchmark problem, and computed sensitivity derivatives are validated by a finite differences study on a generic 2D stenosed geometry. 
        The application included addresses a 3D ducted geometry which features typical characteristics of biomedical devices. An optimized shape, leading to a potential improvement in hemolysis induction up to 22\%, is identified. It is shown, that the improvement persists for different, literature-reported hemolysis-evaluation parameters.
        
        \vspace{0.7cm}

\end{titlepage}
\section{Introduction}

The ever-growing advances in medicine, engineering and material science have led to the development of biomedical devices, such as blood pumps, which allow long-term patient care and significantly improve quality of life. Despite all the advances, a critical task for the design and development of such devices \cite{Thamsen,Yu_opt} is still the minimization of shear-induced blood damage (i.e. hemolysis) to guarantee good bio-compatibility.

\par In the context of this manuscript, hemolysis refers to the mechanical damage of red blood cells due to excessively high stress induced by peculiarities of the blood flow. It can lead to hemoglobinemia, which plays a significant role in the pathogenesis of sepsis, and to increased risk of infection due to its inhibitory effects on the innate immune system \cite{Hemo_sepsis}. Hemolysis induction is encountered in many biomedical devices, where large velocity gradients are found \cite{DeSomer}, as well as in vivo conditions when vessels delivering blood are kinked or stenosed \cite{Goubergrits}. The shape of the respective artificial devices or vessels is believed to play a crucial part in the induction of blood damage due to its decisive fluid dynamic role. A computationally efficient shape optimization framework would be therefore appreciated for the development of next-generation biomedical machinery.

\par 
The success of a numerical optimization process partially depends on the accuracy of the hemolysis-prediction model. While the study of shear-induced hemolysis has been of interest for many experimental,
in vitro conditions for quite some time \cite{Giersiepen, ho, zt}, it is also becoming increasingly important in silico conditions \cite{Yu,FDA}. Computational fluid dynamics (CFD) offer the possibility to predict blood damage in a purely numerical manner and employ the damage prediction model as a cost function in an optimization framework. A variety of such damage prediction models have been proposed \cite{Yu}. Most of them relate hemolysis with the magnitude of the shear stress and an exposure period using a variation of  a power-law formulation, initially suggested by Giersiepen et al. \cite{Giersiepen}.

\par 
Despite the remarkable progress, a numerical model able to satisfactorily predict blood damage in a variety of flows has not yet been established \cite{Goubergrits}. The present manuscript does not aim to advocate the merits of a specific hemolysis evaluation or hemolysis prediction model, but is mainly concerned with the adaptation and integration of a classical model into an adjoint shape optimization framework. Employing an Eulerian hemolysis-prediction model alongside a common CFD solver, we formulate a  Partial-Differential-Equation (PDE) constrained optimization problem, which targets to minimize flow-induced hemolysis by improving the shape. In the context of this paper, hemolysis will be referred to as \textit{objective} while the shape as \textit{control}. In CFD-based optimization, multiple ways, ranging from stochastic \cite{Back_evol,Michalewicz_evol.} to deterministic \cite{Thevenin_grad.,Bertsekas_grad.} optimization methods, could be followed. The present research is concerned with the efficient computation of the derivative of the objective with respect to (w.r.t) the control. The aforementioned derivative 
is subsequently used by a deterministic gradient-based steepest descent method, which drives the controlled shape towards an improved state. To that extent, the continuous adjoint method 
is studied. The adjoint method has been, increasingly, receiving 
attention in terms of CFD-based optimization \cite{papoutsis_adj, papoutsis_adj2, papadimitriou_adj,stuck2013adjoint,Heners} since the pioneering works of Pironneau \cite{pironneau} and Jameson \cite{jameson}, due to its superior computational efficiency. In specific, the attractiveness of the method lies on the fact that the computation of the derivative is independent from the size of the control. For further details on the merits and drawbacks of the adjoint approach, the interested reader is referred to \cite{martins,giles2000introduction}. 

\par
The remainder of this paper is organised as follows: Section \ref{sec:Problem_Formulation} presents the mathematical model of the 
primal and adjoint problem. The same section also reports on the employed numerical procedure and utilized boundary conditions (BCs). In Section \ref{sec:Verification_Studies}, a benchmark case is studied to verify the code reliability. Moreover, the accuracy of the computed sensitivity derivative is assessed in the context of a Finite Differences (FD) study. Subsequently, in Section \ref{sec:application}, the model is applied on a 3D geometry. Results for a full shape optimization process are presented and their dependency on parameters of the nonlinear hemolysis evaluation model is discussed. The paper closes with conclusions and outlines further research in Section \ref{sec:Conclusions}.

Within this publication, Einstein’s summation convention is used for repeated lower-case Latin subscripts. Vectors and tensors are defined with reference to Cartesian coordinates.

\section{Mathematical Model}\label{sec:Problem_Formulation} 

This section is dedicated to the formulation of the primal (physical) and adjoint (dual) problem. The coupling of the hemolysis-prediction model with the primal flow equations is discussed. A presentation of the newly developed adjoint model then follows with detailed discussions on adjoint-hemolysis specific points of interest.

\subsection{Primal Flow}\label{subsec:Primal}
Throughout this paper, blood is treated as an incompressible, Newtonian fluid. The assumption of non-Newtonian behaviour has been shown to satisfactorily predict the flow of blood in tubes with diameter ranging from $130$ to $1000$ $\mu m$ \cite{Merrill}. 
However, all the applications considered herein refer to ducted systems with a larger diameter and thus the Newtonian assumption is preferred. Furthermore, we assume all applications in laminar and steady-state conditions. The blood flow in a domain ($\Omega$) follows from  the Navier-Stokes (NS) equations for the conservation of volume and momentum, viz.
\begin{align}
    R^p   &= - \frac{\partial u_j}{\partial x_j}= 0  \, ,  
    \label{eq:primal_continuity} \\
    R_i^u &=\rho u_j \frac{\partial u_i}{\partial x_j} - \frac{\partial}{\partial x_j}\left[2\mu S_{ij} - p \delta_{ij} \right] = 0 \, . 
    \label{eq:primal_momentum}
\end{align}
Here $u_i$, $p$, $S_{ij}= 0.5(\partial u_i/ \partial x_j +\partial u_j/ \partial x_i)$, $\delta_{ij}$, $\rho$ and $\mu$ refer to the components of the fluid velocity vector, static pressure, components of the strain-rate tensor, Kronecker delta components as well as the fluid density and dynamic viscosity, respectively.

\par In the framework of CFD-based hemolysis analysis, a typical approach is to utilize a prediction or evaluation model in a one-way coupling with the CFD solver \cite{Yu}. This implies that the hemolysis model receives information from the NS equations (\ref{eq:primal_continuity}, \ref{eq:primal_momentum}) with no retro-action on the employed fluid properties ($\rho, \mu$).
The hemolysis prediction model considered in this study originates from the power-law equation, first introduced by Giersiepen et al. \cite{Giersiepen}, and reads
\begin{equation}
    H(\overline{\tau}, t) = C \overline{\tau}^\alpha t^\beta \label{eq:original_power_law} .
\end{equation}
The hemolysis index $H$ denotes a measure for the released hemoglobin to the total hemoglobin within the red blood cell. It is governed by a scalar stress representation $\overline{\tau}$ 
and an exposure time $t$, which represents the duration on which the red blood cell is exposed to the stress. The remaining constants ($C$, $\alpha$, $\beta$) are introduced to fit experimental data. Table \ref{tab:set_of_parameters} summarizes the parameter sets utilized in this study.
\begin{table}[ht]
\begin{tabular}[ht]{cccccr}
	\toprule
 Notation                  & Values $(C, \alpha, \beta)$    & Type of Blood     &  Max. Stress (Pa)\\ \midrule
GW   	                  &  $(3.6 \cdot 10^{-7}$, $2.416$, $0.785)$ 		  & Human    & 255   	    \\
HO   	                  &  $(1.8 \cdot 10^{-8}$, $1.991$, $0.765)$ 	      & Porcine    & 600   	    \\
ZT   	                  &  $(1.2 \cdot 10^{-7}$, $1.992$, $0.661)$ 		  & Ovine    & 320   	     \\ \bottomrule
\end{tabular}
\centering
\caption{Parameter sets of Eqn. (\ref{eq:original_power_law}). The notation corresponds to the initials of the investigators that performed the experiments (GW \cite{Giersiepen}, HO \cite{ho}, ZT \cite{zt}). The final two columns correspond to the type of blood that was used and to the maximum stress that was applied during the experiments.}
\centering
\label{tab:set_of_parameters}
\end{table}
The second invariant of the stress tensor $\tau_{ij}=2\mu S_{ij}$
is frequently used to determine the scalar stress representation $\overline{\tau}$ in combination with a user specific parameter $k=1,2,3$ 
\begin{align}
 \overline{\tau} = \sqrt{-k I_{\tau_2}}
    \qquad \mathrm{with} \qquad 
    I_{\tau_2} = \frac{1}{2}\left[\big(tr(\tau_{ij})^2 - tr(\tau_{ij}^2)\big)\right] 
    \label{eq:scalar_stress}.
\end{align}
Due to $tr(\tau_{ij})  = tr(2\mu S_{ij}) \sim \partial u_i/\partial x_i$, the first contribution to $I_{\tau_2}$ in (\ref{eq:scalar_stress}) vanishes for incompressible Newtonian fluids, which yields
\begin{align}
         \overline{\tau}
                    = \left( 2k \mu^2 S_{ij}S_{ji} \right)^\frac{1}{2} := {\tau^{*}}^\frac{1}{2} .
                    \label{eq:taustar}
\end{align}
To incorporate Eqn. (\ref{eq:original_power_law}) into an adjoint optimization framework, a PDE-based hemolysis prediction is advantageous. To this end, a linearization strategy is used \cite{Garon} for (\ref{eq:original_power_law}), i.e. 
\begin{equation}
    H^{\frac{1}{\beta}} = \left( C^\frac{1}{\beta} \overline{\tau}^{\frac{\alpha}{\beta}} \right) \; t \label{eq:linearize_h}.
\end{equation}
Substituting $H_L= H^{\frac{1}{\beta}}$, the sought material derivative expression follows from (\ref{eq:linearize_h})
\begin{equation}
   \rho \, \frac{\partial H_L}{\partial t}
   + \rho \, u_j \frac{\partial H_L}{\partial x_j} = \rho \, C^{\frac{1}{\beta}} \overline{\tau}^{\frac{\alpha}{\beta}} \, . 
   \label{eq:linearize_h1}
\end{equation}
Assuming steady-state, a modification of the residual form of (\ref{eq:linearize_h1}) reads
\begin{equation}
    R^H = \rho \, u_j \frac{\partial H_L}{\partial x_j} - \rho \, C^{\frac{1}{\beta}} \overline{\tau}^{\frac{\alpha}{\beta}} (1 - H_L) = 0 \quad \text{in  } \Omega \label{eq:model_1}.
\end{equation}
Note that (\ref{eq:model_1}) is enhanced by $(1-H_L)$ to ensure $H_L < 1$. Equation (\ref{eq:model_1}) will be referred to as (primal) hemolysis equation throughout this manuscript and serves to formulate the objective functional of the optimization.
The merits of expression (\ref{eq:model_1}) refer to its straightforward derivation from the original power-law formulation 
(\ref{eq:original_power_law}) and its suitability for showcasing the adjoint method in the context of blood damage. While it is true that hemolysis does not occur for small values of $\overline{\tau} < \overline{\tau}_{th}$, where $\overline{\tau}_{th}$ corresponds to some threshold value (which might also be related to the exposure time), this aspect is deliberately ignored in the present effort. 
Furthermore, the footing of the hemolysis model assumes a homogeneous stress representation. Even though this is a strong assumption for most practical applications, the predictions have been shown to be sufficiently reliable when relatively comparing different geometries \cite{Taskin}. In the present context of shape optimization, it is thus fair to assume that the qualitative information on a reduction of the induced hemolysis is reliable.
\par Equation (\ref{eq:model_1}) serves for the formulation of an adjoint complement in Sec. \ref{subsubsec:Derivation}. Furthermore, the optimization requires a scalar objective functional. An integral index signifying the level of flow-induced hemolysis can be computed from the domain integral  $\int \nabla_j (\rho u_j H_L) \, d\Omega $, or a corresponding boundary integral. Assuming zero inflow fluxes of $H_L$, no-slip velocity on the walls and solenoidal flow fields, only the outlet fluxes remain, which are usually normalized by
the respective mass flux
\begin{equation}
    H_{index} = \frac{\int_{\Gamma_{out}} H \rho u_j n_j d\Gamma}{\int_{\Gamma_{out}} \rho u_j n_j d\Gamma} = \frac{\int_{\Gamma_{out}} H_L^\beta \rho u_j n_j d\Gamma}{\int_{\Gamma_{out}} \rho u_j n_j d\Gamma} \label{eq:objective_normalized}.
\end{equation}
Note that the denominator of Eqn. (\ref{eq:objective_normalized}) is
constant in steady flows of incompressible fluids. Hence, the numerator of (\ref{eq:objective_normalized}) is considered as an appropriate objective or cost-functional in this manuscript.
\subsection{Adjoint Method}\label{subsec:Adjoint}
This subsection outlines the formulation of the adjoint problem 
at hand which is initially expressed as an optimization problem constrained by a set of PDE, namely $R(y,c)=0$, viz. 
\begin{equation}
    \underset{c \in C_{ad}}{\mathrm{min}} J(y,c)  \quad \text{s.t. } R(y,c)=0 \label{eq:optimal_control}.
\end{equation}
Here $J$ is the objective or cost functional, $y$ is the vector of primal state variables, which consists of $p$, $u_i$ and $H_L$, and $c$ is the control parameter which we need to find in the set of admissible states $C_{ad}$. In this study, $c$ refers to the shape of the structure in which blood flows. It is thus a subset of $\partial \Omega$ or $\Gamma$ -the latter notation is preferred throughout this manuscript- and can be further restricted by considering only specific sections of the structure, namely $\Gamma_D \subset \Gamma$, where the index $D$ denotes design.

\par Based on (\ref{eq:objective_normalized}) the objective functional under investigation can be written as
\begin{equation}
    J = 
    \int_{\Gamma_{out}}  H_L^\beta \rho u_i n_i d\Gamma \label{eq:objective_adjoint}=
    \int_{\Gamma_{out}} j_\Gamma d\Gamma ,
\end{equation}
while the set of PDE, $R(y,c)=0$, consists of Eqns. (\ref{eq:primal_continuity}), (\ref{eq:primal_momentum}) and (\ref{eq:model_1}). Having formulated the optimization problem as a constrained problem in Eqn. (\ref{eq:optimal_control}), the Lagrange principle to eliminate the constraints by employing appropriate Lagrange multipliers is utilized \cite{Soto2004OnTC}. In this context, a continuous Lagrange functional is defined as
\begin{equation}
    \pazocal{L} := J + \int_\Omega \left[\hat{p}R^p + \hat{u}_i R^u_i + \hat{h} R^H  \right] d\Omega \label{eq:Lagrangian},
\end{equation}
where $\hat{p}$, $\hat{u}_i$ and $\hat{h}$ are Lagrange multipliers that are referred to as adjoint pressure, components of the adjoint velocity vector and adjoint hemolysis, respectively. Their dimensions $[\hat{u}_\mathrm{i}] = [J]/( [R_\mathrm{i}^u] \, m^3)$, $[\hat{p}] = [J] / ([R^p] \, m^3)$ and $[\hat{h}] =[J]/ ([R^H] m^3)$ depend on the underlying cost functional, where $[J] = ([j_\mathrm{\Gamma}] \, m^2)$ represents the units of the integral boundary-based objective. Since, from a physical point of view, equations $R^p$, $R^u_i$ and $R^H$ are required to be zero, it is apparent that $\pazocal{L}$ is equal to $J$. The optimization problem (\ref{eq:optimal_control}) is, thus, equivalent to ${min} \pazocal{L}(\hat{y},y,c)$, where $\hat{y}$ is the vector of adjoint variables and $y$ is no longer constrained. For a minimum of 
(\ref{eq:Lagrangian}) the total variation $\delta \pazocal{L}$ vanishes
\begin{equation}
    \delta \pazocal{L} = \underbrace{\delta_{\hat{y}} \pazocal{L} \cdot \delta \hat{y}}_{\text{FD1}} + \underbrace{\delta_{y} \pazocal{L} \cdot \delta y}_{\text{FD2}} + \underbrace{\delta_{c} \pazocal{L} \cdot \delta c}_{\text{FD3}} = 0 \label{eq:total_variation_lagrangian} .
\end{equation}
The terms FD1, FD2 and FD3 denote the variation of the Lagrangian 
in the direction of adjoint ($\delta \hat{y}$) and primal ($\delta y$) state as well as the control ($\delta c$), respectively. They can be computed by utilising the functional derivative into the respective direction. FD1 leads to the known set of primal equations
that need to be satisfied for every control state. FD2 yields the accompanying adjoint equations. Finally, FD3 gives rise to a
sensitivity derivative that offers information of the objective w.r.t. the control. Eqn. (\ref{eq:total_variation_lagrangian}) is only satisfied when a global or local minimum is reached. Hence, the following three optimality conditions are obtained for vanishing FD2
\begin{align}
    \delta_p \pazocal{L} \cdot \delta p = 0 \quad \forall \delta p 
    \, , \qquad 
    \delta_{u_i} \pazocal{L} \cdot \delta u_i = 0 \quad \forall \delta u_i 
    \, , \qquad 
    \delta_{H_L} \pazocal{L} \cdot \delta H_L = 0 \quad \forall \delta H_L \label{eq:optimality_hemolysis}.
\end{align}
A detailed overview on PDE constraint optimization can be found in \cite{hinze2008optimization}.

\subsubsection{Adjoint Flow Equations} \label{subsubsec:Derivation}
Each optimality condition (\ref{eq:optimality_hemolysis})
leads to a PDE that needs to be satisfied in order for FD2 to vanish. The essence of the adjoint method is to identify an adjoint state $\hat{y}$ that satisfies (\ref{eq:optimality_hemolysis}),
so that the total variation $\delta \pazocal{L}$ depends only on the variation of the control ($\delta_{c} \pazocal{L} \cdot \delta c$).
For the purpose of this paper it is deemed beneficial to split (\ref{eq:Lagrangian}) into sub-integrals consisting of isolated terms of the primal equations, viz.
\begin{align}
\underbrace{\int_\Omega \hat{p}\Big(-\frac{\partial u_j}{\partial x_j}\Big) d\Omega}_{I_1} &\; + \; 
 \bigg[
  \underbrace{\int_\Omega \hat{u}_i \Big(\rho u_j\frac{\partial u_i}{\partial x_j}\Big) d\Omega}_{I_2}  
 + \underbrace{\int_\Omega \hat{u}_i\Big(-\frac{\partial}{\partial x_j}(2\mu S_{ij})\Big) d\Omega}_{I_3} 
 + \underbrace{\int_\Omega \hat{u}_i\Big(\frac{\partial p}{\partial x_i}\Big) d\Omega}_{I_4}  \,
 \bigg] 
 \nonumber \\
 &\; + \; 
  \bigg[\underbrace{\int_\Omega \hat{h}\Big(u_j \frac{\partial H_L}{\partial x_j}\Big) d\Omega}_{I_5}
  + \underbrace{\int_\Omega \hat{h}\Big(- C^{\frac{1}{\beta}} \overline{\tau}^{\frac{\alpha}{\beta}} (1 - H_L) \Big) d\Omega}_{I_6}\bigg].
\nonumber
\end{align}
The computation of the functional derivatives (\ref{eq:optimality_hemolysis}) requires the computation of the derivatives of each sub-integral $I_k$, with $k=1,2...,6$, as well as the derivatives of the objective functional (\ref{eq:objective_adjoint}). The latter involves $\delta_{p} J \cdot \delta p  =  0$ and 
\begin{align}
     \delta_{u_i} J \cdot \delta u_i =  \int_{\Gamma_{out}} \delta u_i \Big(\rho H_L^\beta n_i \Big) d\Gamma  
     \, , \quad 
     \delta_{H_L} J \cdot \delta H_L =  \int_{\Gamma_{out}} \delta H_L \Big(\rho \beta H_L^{(\beta-1)} u_i n_i \Big) d\Gamma. \label{eq:objective_var_H}
\end{align}
Note that the objective functional and its directional derivatives only exist at the outlet of the domain, which is not a design surface ($\Gamma_D$). The contribution of the objective functional to the adjoint equations thus appears only in the outlet boundary conditions. In an analogy to primal duct flows which are driven by a difference between the inlet and the outlet pressure, the dual flow is driven by the relative difference of the hemolysis index.

The calculation of the variations $\delta_y I_k \cdot \delta y$, with $k=1,2..,4$ has been previously shown in many manuscripts \cite{Soto2004OnTC,papoutsis_adj}. It is therefore omitted from the main body of this paper and can be found in App. A.
The derivation for the hemolysis relevant terms $I_5$ and $I_6$ represents a core contribution of this paper.
For $\delta_y I_5 \cdot \delta y$, we can utilize Gauss's divergence theorem and obtain 
\begin{align}
    \delta_p I_5 \cdot \delta p &= 0 \label{eq:I5_p_var}\\
    \delta_{u_i} I_5 \cdot \delta u_i &= \int_\Omega \delta u_i \Big( \hat{h} \frac{\partial H_L}{\partial x_i}\Big) \, d\Omega = \int_\Gamma (\delta u_i n_i) H_L \hat{h} \, d\Gamma - \int_\Omega \delta u_i \Big(H_L \frac{\partial \hat{h}}{\partial x_i}\Big) \, d\Omega \label{eq:I5_u_var} \\
    \delta_{H_L} I_5 \cdot \delta H_L &= \int_\Omega \hat{h} u_j \frac{\partial \delta H_L}{\partial x_j} \, d \Omega = \int_\Gamma \delta H_L (u_j n_j) \hat{h} \, d \Gamma - \int_\Omega \delta H_L \Big ( u_j \frac{\partial \hat{h}}{\partial x_j} \Big) \, d\Omega  \label{eq:I5_H_var}.
\end{align}
Although Eqn. (\ref{eq:I5_u_var}) doesn't necessarily require integration by parts to be included in the adjoint equations, the respective volume integral is expanded in the context of this work for computational purposes.

Following the same strategy for $I_6$, yields  $\delta_p I_6 \cdot \delta p = 0$. As for $\delta_{u_i} I_6 \cdot \delta u_i$, we first define $M=C^{\frac{1}{\beta}}(1-H_L)$ as a scalar quantity not subjected to any variation w.r.t the velocity. Using the scalar stress expression (\ref{eq:taustar}), the integral $I_6$ is expressed as $I_6 = - \int_\Omega \hat{h} M {\tau^{*}}^{\frac{\alpha}{2\beta}} d\Omega$ and its linearized variation in the velocity direction reads
\begin{align}
    \delta_{u_i} I_6 \cdot \delta u_i &=-\int_\Omega \hat{h} M \frac{\alpha}{2\beta}{\tau^{*}}^{(\frac{\alpha}{2\beta}-1)}\left( \delta_{u_i} \tau^{*} \cdot \delta u_i \right) d\Omega \nonumber \\
    &=-\int_\Omega \hat{h} \Big( M \frac{\alpha}{2\beta}{\tau^{*}}^{(\frac{\alpha}{2\beta}-1)} 4 k \mu^2  S_{ij} \Big) \delta S_{ij}  \, d\Omega  \label{eq:I6_u_var},
\end{align}
where $\delta S_{ij} = 0.5 \left(\partial \delta u_i/\partial x_j + \partial \delta u_j/\partial x_i \right)$. By defining $B_{ij}= 2 M k \mu^2 \frac{\alpha}{\beta}{\tau^{*}}^{(\frac{\alpha}{2\beta}-1)} S_{ij}$, a compact form of (\ref{eq:I6_u_var}) is further expanded via integration by parts as 
\begin{align}
    \delta_{u_i} I_6 \cdot \delta u_i &= -\int_\Omega \hat{h} \, \frac{B_{ij}}{2}\Big(\frac{\partial \delta u_i}{\partial x_j} + \frac{\partial \delta u_j}{\partial x_i}\Big ) d\Omega 
    = -\int_\Omega \hat{h}B_{ij}\Big(\frac{\partial \delta u_i}{\partial x_j}\Big) d\Omega \nonumber \\
    &= -\int_\Gamma \delta u_i \left( \hat{h} B_{ij} n_j \right) d\Gamma + \int_\Omega \delta u_i \left( \frac{\partial (\hat{h}B_{ij})}{\partial x_j} \right) d\Omega \label{eq:I6_u_var_final}.
\end{align}
Finally, $\delta_{H_L} I_6 \cdot \delta H_L$ is calculated as 
\begin{equation}
    \delta_{H_L} I_6 \cdot \delta H_L = \int_\Omega \delta H_L \Big( \hat{h} C^{\frac{1}{\beta}} \overline{\tau}^{\frac{\alpha}{\beta}} \Big) d\Omega \label{eq:I6_H_var}.
\end{equation}
\par Having expressed the variation of $J$ and $I_k$ in terms of several boundary and volume integrals, we can superpose the variation of the Lagrange functional as the sum of one boundary and volume integral, viz.
\begin{equation}
 \delta_y \pazocal{L} \cdot \delta y = \int_\Gamma \delta y (\cdot) d\Gamma + \int_\Omega \delta y (\cdot) d\Omega \label{eq:lagrange_variation_y}.
\end{equation}
The adjoint equations as well as their corresponding boundary conditions arise by demanding that the integrands in Eqn. (\ref{eq:lagrange_variation_y}) vanish for any test function $\delta y$. Therefore, following the aforementioned methodology for the optimality conditions (\ref{eq:optimality_hemolysis}), the resulting constraints in the interior of the domain ($\Omega$) read
\begin{alignat}{2}
    \hat{R}^p &= -\frac{\partial \hat{u}_i}{\partial x_i} = 0 ,
    \label{eq:adjoint_continuity}\\
    \hat{R}_i^u &= \rho \hat{u}_j \frac{\partial u_j}{\partial x_i} - \rho u_j \frac{\partial \hat{u}_i}{\partial x_j}-\frac{\partial}{\partial x_j}\Big(2 \mu \hat{S}_{ij} - \hat{p}\delta_{ij}\Big) +\frac{\partial (\hat{h}B_{ij})}{\partial x_j} - H_L\frac{\partial \hat{h}}{\partial x_i} = 0 , 
    \label{eq:adjoint_momentum}\\
    \hat{R}^H &= u_j \frac{\partial \hat{h}}{\partial x_j} - C^{\frac{1}{\beta}}\overline{\tau}^{\frac{\alpha}{\beta}} \hat{h} = 0 
    \label{eq:adjoint_hemolysis} .
\end{alignat}
Equations (\ref{eq:adjoint_continuity}) as well as (\ref{eq:adjoint_momentum}) represent the adjoint companions to the continuity and momentum equation, respectively. In this sense, Eqn. (\ref{eq:adjoint_hemolysis}) is referred to as adjoint hemolysis equation throughout this manuscript. The adjoint momentum equation is enhanced with the last two terms on the left-hand-side (LHS) of the equation that include contributions from adjoint and primal hemolysis. This is due to the fact that on the primal side of the system, the velocity appears twice in the hemolysis equation, once in the advection  and once in the form of a gradient in the source term. Both terms act similar to a pressure term in (\ref{eq:adjoint_momentum}) and thereby drive the adjoint flow field.

Furthermore, it is also worth saying that in contrast to its primal counterpart, the adjoint hemolysis equation doesn't require any information from the solution of the adjoint continuity and momentum equations. Once again, we have a one-way coupling but this time the direction of the coupling and thereby the algorithmic order of sequence is reversed. Practically this means that the adjoint hemolysis equation could be solved after the solution of the primal equations and prior to the solution of the other two adjoint Eqns. (\ref{eq:adjoint_continuity}) \& (\ref{eq:adjoint_momentum}).

The adjoint system (\ref{eq:adjoint_continuity}-\ref{eq:adjoint_hemolysis}) is an extension of a classical adjoint (incompressible)  NS system, which corresponds to  (\ref{eq:adjoint_continuity}-\ref{eq:adjoint_momentum}) excluding the two hemolysis terms. This is similar to the adjoint complement of the Volume of Fluid (VoF) method \cite{kroger2018adjoint, kuhl2020adjoint}, used for multi-phase applications, and enables a straightforward implementation, provided that an adjoint multi-phase solver exists.
Although the adjoint equations are per definition linear, several cross coupling terms might introduce a severe stiffness to the densely coupled PDE system. A viable workaround to stabilize the iterative procedure refers to the introduction of additional diffusive terms or adjoint pseudo time-stepping, cf. \cite{giles2010convergence, kroger2018adjoint, kuhl2020adjoint}.

\par The adjoint BCs arise by fulfilling the requirement of eliminating the surface integral of  (\ref{eq:lagrange_variation_y}). By expanding this necessity for every primal state variable the BCs read 
\begin{align}
\hat{u}_i n_i &= 0 \quad \text{along  } \Gamma , \nonumber \\
-\hat{p}n_i + \rho \hat{u}_i u_j n_j + \mu \frac{\partial \hat{u}_i}{\partial x_j} n_j + H_L \hat{h} n_i - \hat{h}B_{ij}n_j &= 0 \quad \text{along  } \Gamma \setminus \Gamma_{out} ,  \nonumber \\
-\hat{p}n_i + \rho \hat{u}_i u_j n_j + \mu \frac{\partial \hat{u}_i}{\partial x_j} n_j + H_L \hat{h} n_i - \hat{h}B_{ij}n_j + \rho H_L^\beta n_i &= 0 \quad \text{along  }  \Gamma_{out} , \nonumber \\ 
-\hat{h}(u_j n_j) &= 0 \quad \text{along  } \Gamma \setminus \Gamma_{out} , \nonumber \\
\hat{h} + \rho \beta H_L^{(\beta-1)}&= 0 \quad \text{along  }  \Gamma_{out} \label{eq:adjoint_hemolysis_BC_2}.
\end{align}

\subsubsection{Computation of Surface Sensitivity} \label{subsubsec:Sensitivity}
Ensuring a vanishing residual of the primal and adjoint system of equations leads to vanishing FD1 and FD2 terms of (\ref{eq:total_variation_lagrangian}). The FD3 term allows for the computation of a surface sensitivity which is used by a gradient-based algorithm to drive the shape towards an improved state. Following (\ref{eq:Lagrangian}), FD3 is expressed as
\begin{equation}
    \delta_c \pazocal{L} \cdot \delta c = \delta_c J \cdot \delta c + \int_\Omega \hat{y} \Big(\delta_c R \cdot \delta c \Big) d\Omega \label{eq:surface_sen_1}.
\end{equation}
Since the primal flow equations (\ref{eq:primal_continuity}, \ref{eq:primal_momentum}, \ref{eq:model_1}) are also fulfilled under the total variation, one obtains
\begin{equation}
    \delta R(y,c) = \delta_c R \cdot \delta c + \delta_y R \cdot \delta y = 0 \label{eq:surface_sen_2}.
\end{equation}
Therefore, the variation w.r.t. the control transforms to
\begin{equation}
 \delta_c R \cdot \delta c = - \delta_y R \cdot \delta y  \label{eq:surface_sen_3}.
\end{equation}
 Furthermore, based on the formulation of our objective functional $J$, which is defined only on the outlet, $\delta_c J \cdot \delta c = 0$ since $c \equiv \Gamma_D$ and $\Gamma_D \cap \Gamma_{out} = \emptyset$. The sensitivity derivative can thus be written as
\begin{equation}
\delta_c \pazocal{L} \cdot \delta c = - \int_\Omega \hat{y} \Big(\delta_y R \cdot \delta y \Big) d\Omega \label{eq:surface_sen_4}.
\end{equation}
The right-hand-side (RHS) of (\ref{eq:surface_sen_4}) is developed further and by taking into consideration the BCs of the problem as well as a Taylor expansion of the velocity w.r.t a perturbation of the design wall (see App. B or \cite{kuhl2020adjoint_2}), the sensitivity to be computed reads
\begin{equation}
    S_L = \int_{\Gamma_{D}} - \left[ \mu \frac{\partial u_{i(t)}}{\partial n}\frac{\partial \hat{u}_{i(t)}}{\partial n}\right] d\Gamma \label{eq:surface_sen_5},
\end{equation}
where $u_{i(t)}$, $ \hat{u}_{i(t)}$ denote the part of the primal and adjoint velocity tangential to the surface, respectively and $n$ denotes the normal to the surface. Interestingly, the shape derivative is not directly affected by the adjoint or primal hemolysis fields. Nevertheless, the primal and adjoint hemolysis parameters drive the adjoint flow field, cf. (\ref{eq:adjoint_momentum}), and thereby propagate into the shape derivative through the adjoint velocity gradient.

\subsection{Boundary Conditions}\label{subsec:BC}

In what concerns the application studied in this work, the flow boundary $\Gamma$ consists of three parts, namely the inlet ($\Gamma_{in}$), outlet ($\Gamma_{out}$) and wall ($\Gamma_{W}$). We split  the wall boundary into two portions, $\Gamma_D$ and $\Gamma_B = \Gamma_{W} \setminus \Gamma_D$, where the first one involves the parts under design while the latter one is bounded to the initial configuration. The complete boundary domain is thus described as $\Gamma = \Gamma_{in} \cup \Gamma_{out} \cup \Gamma_{D} \cup \Gamma_{B}$.

\par While the BCs of the primal system of equations are selected based on the physical properties of the problem, the adjoint BCs are deduced based on the necessity of vanishing the surface integrals of Eqn. (\ref{eq:lagrange_variation_y}). In many boundary patches this necessity is inherently satisfied due to the primal BCs, though in general it reduces to satisfying Eqns. 
(\ref{eq:adjoint_hemolysis_BC_2}). Furthermore, it is worth mentioning that the BCs strongly depend on the objective functional, if that is defined in a subset of $\Gamma$. The complete set of BCs for the primal and adjoint problem are summarized in Tab. \ref{tab:Boundary_Conditions}.
\begin{table}[ht]
\begin{tabular}[ht]{|c|c|c|c|c|c|c|}
\hline
 Boundary Patch                 & $u_i$  & $p$    & $H_L$ & $\hat{u}_i$ & $\hat{p}$ & $\hat{h}$  \\  

 \hline \xrowht[()]{10pt}
 $\Gamma_{in}$	                 & $u_i = u_i^{in}$  &  $\frac{\partial p}{\partial n} = 0$   & $H_L = 0$ &  $\hat{u}_i = 0$ & $\frac{\partial \hat{p}}{\partial n}=0$ 	& $\frac{\partial \hat{h}}{\partial n}=0$   \\
 \hline \xrowht[()]{10pt}
$\Gamma_{out}$  	             &  $\frac{\partial u_i}{\partial n} = 0$	& $\frac{\partial p}{\partial n} = 0$    & $\frac{\partial H_L}{\partial n} = 0$  & $\hat{u}_i = 0$ & 2nd Eqn. (\ref{eq:adjoint_hemolysis_BC_2})	& 5th Eqn. (\ref{eq:adjoint_hemolysis_BC_2})	    \\
 \hline \xrowht[()]{10pt}
$\Gamma_{W}$   	                  &  $u_i = 0$		  & $\frac{\partial p}{\partial n} = 0$   & $\frac{\partial H_L}{\partial n} = 0$ & $\hat{u}_i = 0$  & $\frac{\partial \hat{p}}{\partial n}=0$	& $\frac{\partial \hat{h}}{\partial n}=0$	     \\
 \hline

\end{tabular}
\centering
\caption{Boundary conditions for closing the primal and adjoint equation system. 
}
\centering
\label{tab:Boundary_Conditions}
\end{table}
\FloatBarrier

\subsection{Numerical Procedure}\label{subsec:Numerical_Procedure} 

The numerical procedure for the solution of the primal and adjoint system is based upon the Finite Volume Method (FVM) employed by FreSCo$^+$ \cite{rung2009challenges}. Analogue to the use of integration-by-parts in deriving the continuous adjoint equations \cite{kuhl2019decoupling, kuhl2020continuous}, 
summation-by-parts is employed to derive the building blocks of the discrete adjoint expressions. A detailed derivation of this hybrid adjoint approach can be found in \cite{ stuck2013adjoint, kroger2018adjoint, kuhl2020adjoint}. The last two terms of the adjoint momentum equation, involving hemolysis contributions, are added explicitly to the RHS. The segregated algorithm uses a cell-centered, collocated storage arrangement for all transport properties. The implicit numerical approximation is second order accurate in space and supports polyhedral cells. Both, the primal and adjoint pressure-velocity coupling is based on a SIMPLE method and possible parallelization is realized by means of a domain decomposition approach \cite{yakubov2013hybrid, yakubov2015experience}. 

\subsection{Optimization Framework}\label{subsec:Opt_framework}

\begin{figure}[ht!]
\centering
\usetikzlibrary{shapes.geometric}

\tikzstyle{startstop} = [rectangle, rounded corners, minimum width=4cm, minimum height=1cm, text centered, draw=black, fill=red!50]
\tikzstyle{io} = [trapezium, trapezium left angle=70, trapezium right angle=110, minimum width = 5cm, minimum height=1cm,  text centered, draw=black, text width=4cm, fill=blue!20 ]
\tikzstyle{process} = [rectangle, minimum width = 5cm, minimum height=1cm,  text centered, draw=black, text width=5cm, fill=orange!40]
\tikzstyle{decision} = [diamond, minimum width = 2cm, minimum height=1cm, text width=2cm,  text centered, draw=black, fill=green!30]
\tikzstyle{arrow} = [thick, ->, >=stealth]

\begin{tikzpicture}[node distance=2cm]

\node (start) [startstop] {Start};
\node (in_shape) [io, below of=start, yshift=-1cm] {Initial shape $c^0$, \\Solution settings (fluid properties, numerical schemes, e.t.c)};
\node (primal) [process, below of=in_shape,yshift=-1.0cm ] {Solve primal problem in $c^i$, \\ Eqns. (\ref{eq:primal_continuity}), (\ref{eq:primal_momentum}), (\ref{eq:model_1}) $\xrightarrow{}$ Acquire $y$.};
\node (adjoint) [process, below of=primal,yshift=-1.0cm ] {Solve adjoint problem in $c^i$ based on known $y$, \\ Eqns. (\ref{eq:adjoint_continuity}), (\ref{eq:adjoint_momentum}), (\ref{eq:adjoint_hemolysis}) $\xrightarrow{}$ Acquire $\hat{y}$.};
\node (sensitivity) [process, below of=adjoint,yshift=-1.0cm ] {Compute surface sensitivity based on known $y$ and $\hat{y}$, \\ Eqn. (\ref{eq:surface_sen_5}) $\xrightarrow{}$ Acquire $S_L$.};
\node (gradient) [process, below of=sensitivity,yshift=-1.0cm ] {Update shape to $c^{i+1}$ using the steepest descent method.};
\node (convergence) [decision, right of=adjoint,xshift=3cm, yshift=-0.7cm ] {Is $i<i_{max}$?};
\node (end) [startstop, right of = convergence,xshift=3cm ] {End};

\draw [arrow] (start) -- (in_shape);
\draw [arrow] (in_shape) -- (primal);
\draw [arrow] (primal) -- (adjoint);
\draw [arrow] (adjoint) -- (sensitivity);
\draw [arrow] (sensitivity) -- (gradient);
\draw [arrow] (gradient) -| (convergence);
\draw [arrow] (convergence) |-node[anchor=west, yshift=-1.5cm]{YES} (primal);
\draw [arrow] (convergence) --node[anchor=south,]{NO}(end);

\end{tikzpicture}
\caption{Flowchart of adjoint shape optimization framework for one design candidate $i$.
}
\label{fig:flowchart}
\end{figure}
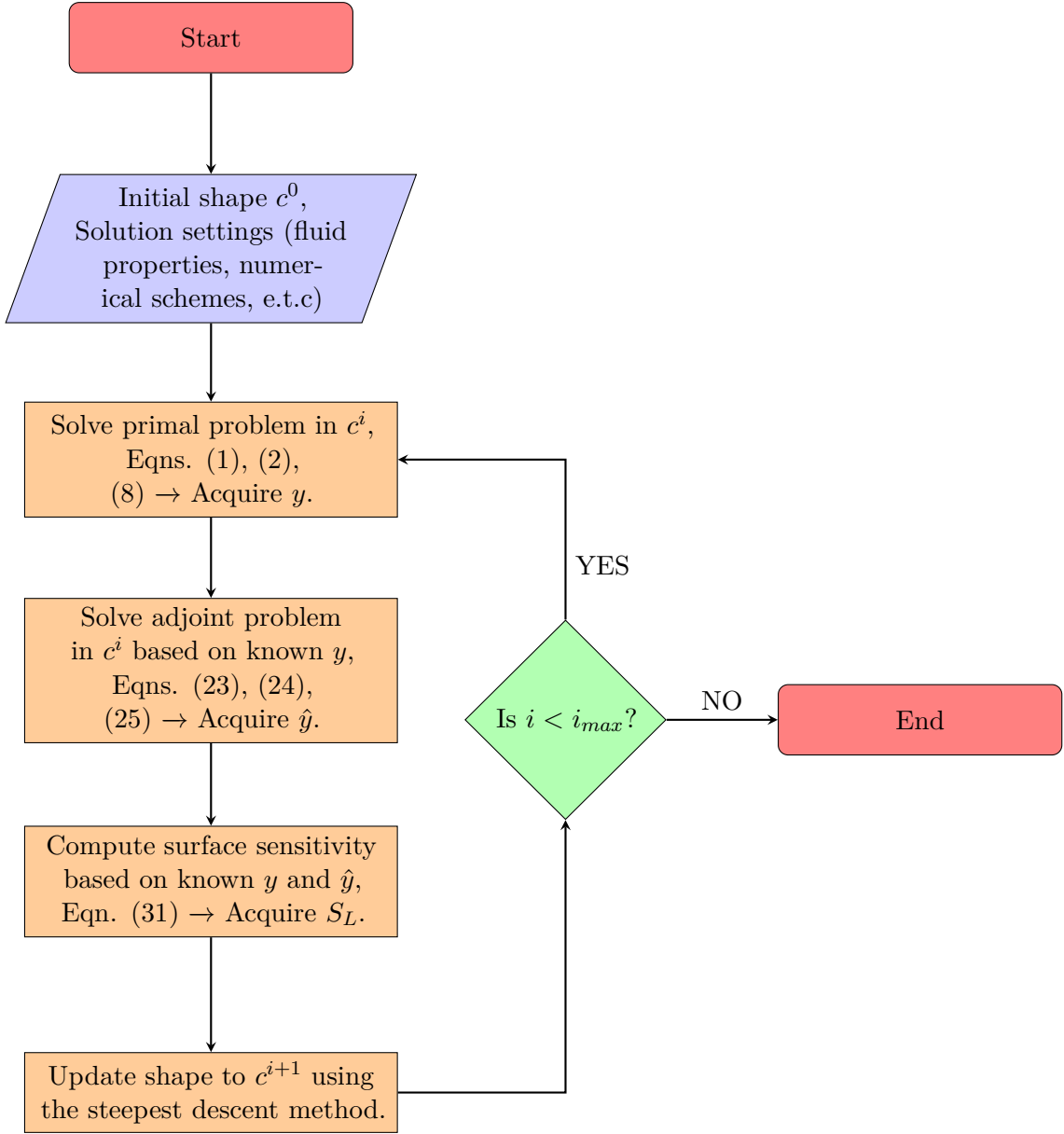

The complete shape-optimization framework is summarized in Fig. \ref{fig:flowchart}. The efficiency of the method is illustrated by taking into consideration the required computational budget for one complete optimization process based on the presented flowchart. Assuming that the cost for solving the primal problem is equal to 1 equivalent flow solution (EFS) then the cost for solving the adjoint one is approximately also equal to 1 EFS. The final two processes (represented by rectangles on the flowchart) have a practically negligible cost compared to the EFS unit. The complete optimization cost is thus equal to $i_{max}$ times 2 EFS, regardless of the size of the control which for our case is proportional to the number of discretized nodes on $\Gamma_D$. In the context of this study, the steepest descent method is considered to advance the shape.

The computed surface sensitivity $S_L$ might possibly be rough. We thus employ the Laplace-Beltrami \cite{kroger2015cad} metric to extract a smooth gradient $G_{L_i}$ through a numerical approximation of 
 \begin{equation}
     G_{L_i} - \lambda \Delta_{\Gamma} G_{L_i} = S_L n_i \label{eq:gradient},
 \end{equation}
where $\Delta_{\Gamma}$ refers to the Laplace-Beltrami operator ($\Delta_{\Gamma} = \Delta - \Delta_\mathrm{n}$), $\lambda$ corresponds to a user-defined  control of smoothing and $n_i$ denotes the normal vector at each face of $\Gamma$. 
 Once the smooth gradient field is available, the 
 displacement field $d_i$ is computed from Eqn. 
 (\ref{eq:displ_gamma_b}),
 \begin{align}
     \frac{\partial}{\partial x_j}\left[ q \frac{\partial d_i}{\partial x_j}\right] &= 0 \quad \text{in   } \Omega , \qquad 
     d_i &= G_{L_i} \quad \text{along     } \Gamma_D , \qquad 
     d_i &= 0      \quad \quad \text{along     }  \Gamma \setminus \Gamma_D . \label{eq:displ_gamma_b}
 \end{align}
 where $q = 1/(W_D + \epsilon)$, $W_D$ is the wall normal distance and $\epsilon=10^{-20}$.

 \FloatBarrier
\section{Verification and Validation Studies}\label{sec:Verification_Studies}
This section verifies the implementation of the primal and adjoint hemolysis system for a benchmark problem, as suggested
by the Food \& Drug Administration (FDA) \cite{FDA_Verification}.  
Analytical solutions are derived and compared with numerical predictions. Subsequently, a finite difference study is conducted 
on a 2D geometry to validate the sensitivity.  

\subsection{Verification}\label{subsec:Analytical_Solutions}
The benchmark problem refers to a fully developed pipe flow, considered on a 3D mesh. For brevity reasons, the derivation of the analytical primal hemolysis solution is skipped and the reader is referred to \cite{FDA_Verification}. The solution of the primal flow reads
\begin{align}
    u_z(r) = U_{max} \left(1- \left(\frac{r}{R}\right)^2\right),  \qquad
     H(r,z) = C \frac{\overline{\tau}^\alpha z^\beta}{u_z^\beta}, \qquad 
     \overline{\tau}(r)  = 2 \mu U_{max} \frac{r}{R^2}  
    \label{eq:pipe_fully_dev_vel}, 
\end{align}
where $r$ corresponds to the radial direction, $R$ marks the pipe radius, $U_{max}$ refers to the centerline velocity, $z$ to the axial coordinate and the direction of the primary flow velocity $u_z$. 

\begin{figure}[ht!]
\begin{minipage}{.5\textwidth}
\centering
\includegraphics[trim=0.8cm 0.6cm 0.8cm 0.6cm, width=1.0\textwidth]{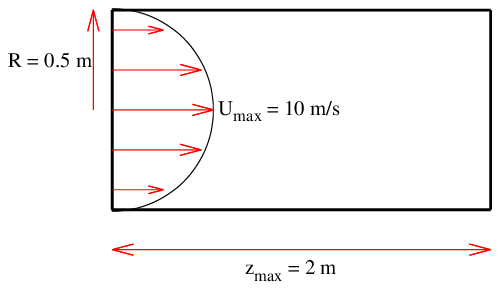}
\caption{Sketch of the pipe flow benchmark.\\ \label{fig:pipe_sketch}}
\end{minipage}%
\hspace{5mm}
\begin{minipage}{.5\textwidth}
\centering
\includegraphics[width=0.65\textwidth]{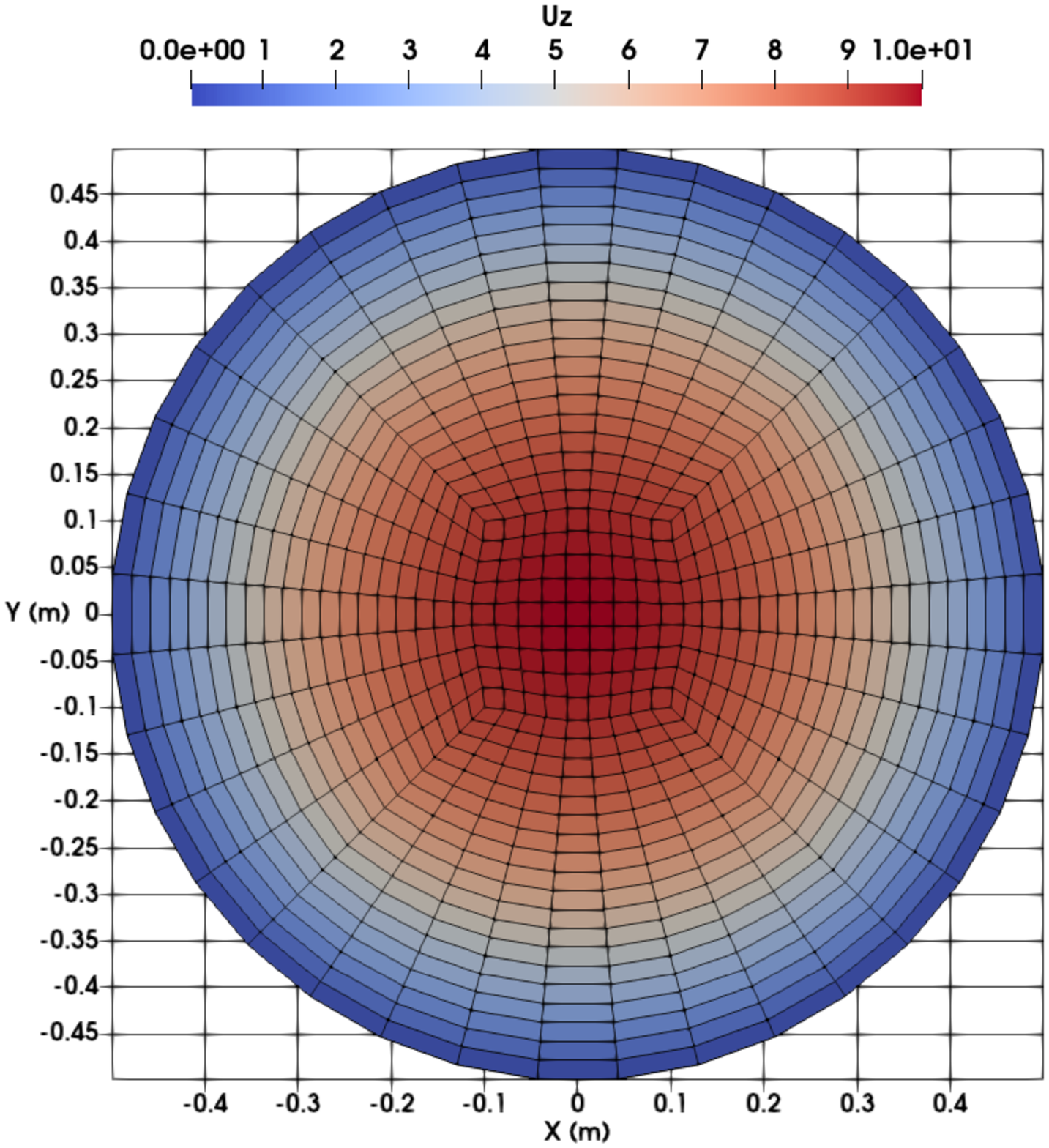}
\caption{Cross-section of the pipe mesh coloured with the axial velocity.
\label{fig:pipe_grid} }
\end{minipage}
\end{figure}

The considered verification case is sketched in Fig. \ref{fig:pipe_sketch}. The entrance and exit planes of the pipe refer to $z=0$ and $z_{max}=2$ $m$. The pipe radius reads $R=0.5$ $m$. The fully developed axial velocity profile of the laminar flow 
utilizes $U_{max}=10$ $m/s$. The fluid properties are selected based on a prescribed Reynolds number of $\mathrm{Re} = 2U R / \nu = 2000$, where $U$ and $\nu$ refer to the bulk velocity and kinematic viscosity, respectively. The three-dimensional geometry is discretized with approximately $75$k control volumes on a structured grid. A cross-section of the computational grid employed is shown in Fig. \ref{fig:pipe_grid}.

Due to the decoupled nature of the adjoint hemolysis equation (\ref{eq:adjoint_hemolysis}),  
an analytical solution can be stated using the analytical primal flow solution (\ref{eq:pipe_fully_dev_vel}), viz.

\begin{equation}
  u_z \frac{d \hat{h}}{d z} - C^{\frac{1}{\beta}} \overline{\tau}^{\frac{\alpha}{\beta}} \hat{h} = 0. \label{eq:adjoint_analytical_2}
\end{equation}
If we abbreviate $\Lambda = C^{\frac{1}{\beta}}\overline{\tau}^{\frac{\alpha}{\beta}} $, then the solution of (\ref{eq:adjoint_analytical_2}) reads
\begin{equation}
    \hat{h} = K \mathrm{e}^{\frac{\Lambda}{u_z} z} \label{eq:adjoint_analytical_3},
\end{equation}
 The integration constant $K$ is calculated based on the BCs of the adjoint hemolysis equation which results from satisfying the final expression in (\ref{eq:adjoint_hemolysis_BC_2}) and demands $\hat{h}|_{out} = - \rho \beta H_L^{\beta-1}|_{out}$. Finally, the adjoint hemolysis field in a fully developed pipe flow reads
\begin{equation}
        \hat{h} = - \rho \beta H_L^{\beta - 1}\Big|_{out} \mathrm{e}^{\frac{\Lambda}{u_z} (z - z_{max})}
    \label{eq:adjoint_analytical_final},
\end{equation}
where $H_L|_{out}$ is calculated by setting $z=z_{max}$ in (\ref{eq:pipe_fully_dev_vel}) together with  $H_L^\beta = H$. 

\begin{figure}[ht!]
\begin{minipage}{.5\textwidth}
\centering
\includegraphics[width=1.0\textwidth]{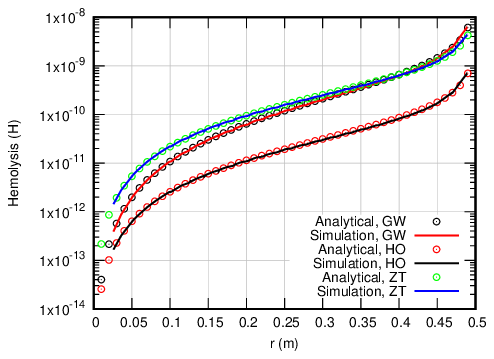}
\caption{Primal hemolysis profiles at $z=1.0$ $m$ for three different parameter sets ($C$, $\alpha$, $\beta$). The notation follows from Tab. \ref{tab:set_of_parameters}. Continuous lines correspond to computations while symbols to analytical values.  \label{fig:pipe_h}}
\end{minipage}%
\hspace{5mm}
\begin{minipage}{.5\textwidth}
\centering
\includegraphics[width=1.0\textwidth]{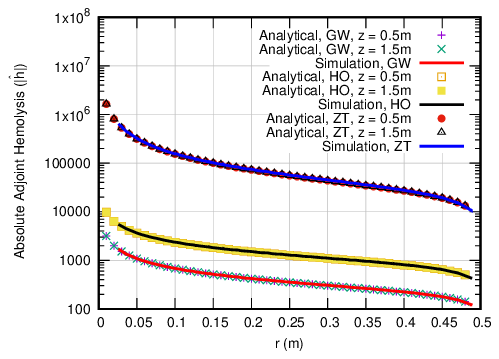}
\caption{Magnitude of adjoint hemolysis profiles for three different  parameter sets ($C$, $\alpha$, $\beta$). 
Notation as in Fig. \ref{fig:pipe_h}. Analytical results are reported for two longitudinal positions, $z=0.5$ and $z=1.5$ $m$. \\ \label{fig:adj_pipe_h} }
\end{minipage}
\end{figure}

Figure \ref{fig:pipe_h}  compares the computed primal hemolysis solution against analytical values as calculated by Eqn. (\ref{eq:pipe_fully_dev_vel}). The comparison is realized for all three sets of hemolysis model parameters outlined in Tab. \ref{tab:set_of_parameters}. As can be seen, all computed values are fitting the analytical solutions to a satisfying degree. 
Figure \ref{fig:adj_pipe_h}
compares the computed adjoint hemolysis field
against the analytical solutions (\ref{eq:adjoint_analytical_final}). Due to the nature of the BCs of the adjoint hemolysis equation, the field values are negative and thus the absolute values are presented to support a log-scale of the ordinate. Predictions are nearly identical to  
analytical solutions for all three set of parameters. Both the primal and adjoint hemolysis solvers
are thus considered verified.

We would like to remark on the nature of the adjoint hemolysis equation. As can be seen in Fig. \ref{fig:adj_pipe_h}, the adjoint hemolysis profile changes only slightly for different axial positions. The BC at the outlet, which reads $\hat{h}|_{out}=-\rho \beta H_L^{\beta-1}|_{out}$, dominates the complete field since all three cases correspond to $\beta < 1$ and $H_L<<1$. To avoid numerical errors that would arise for $H_L|_{out}=0$, the BCs on the outlet is reformulated to $\hat{h}|_{out}=-\rho \beta (H_L+\epsilon)^{(\beta-1)}|_{out}$, where $\epsilon=10^{-20}$. Based on the previous comments, one could assume that applying the bulk outlet adjoint hemolysis profile to the whole field suffices. However, due to the existence of the final two terms in the adjoint momentum, (cf. Eqn.  (\ref{eq:adjoint_momentum})), this assumption would fall short on capturing the conceptual description of the complete model.  

\FloatBarrier

\subsection{Sensitivity Validation}\label{subsec:FD}
\begin{figure}[ht!]
\begin{minipage}{.5\textwidth}
\centering
\includegraphics[trim=0.8cm 1.2cm 0.8cm 1.1cm, width=1.0\textwidth]{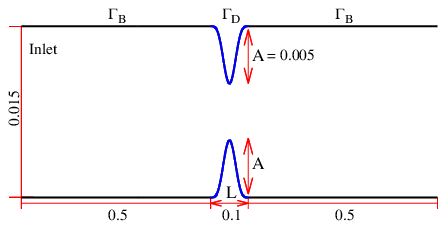}
\caption{
Geometry for the 2D FD study. Blue lines correspond to the design walls ($\Gamma_D$) while black lines indicate bounded walls ($\Gamma_B$). All dimensions in (m). (Figure not in scale). \label{fig:FD_geometry}}
\end{minipage}
\hspace{5mm}
\begin{minipage}{.5\textwidth}
\centering
\includegraphics[width=0.8\textwidth]{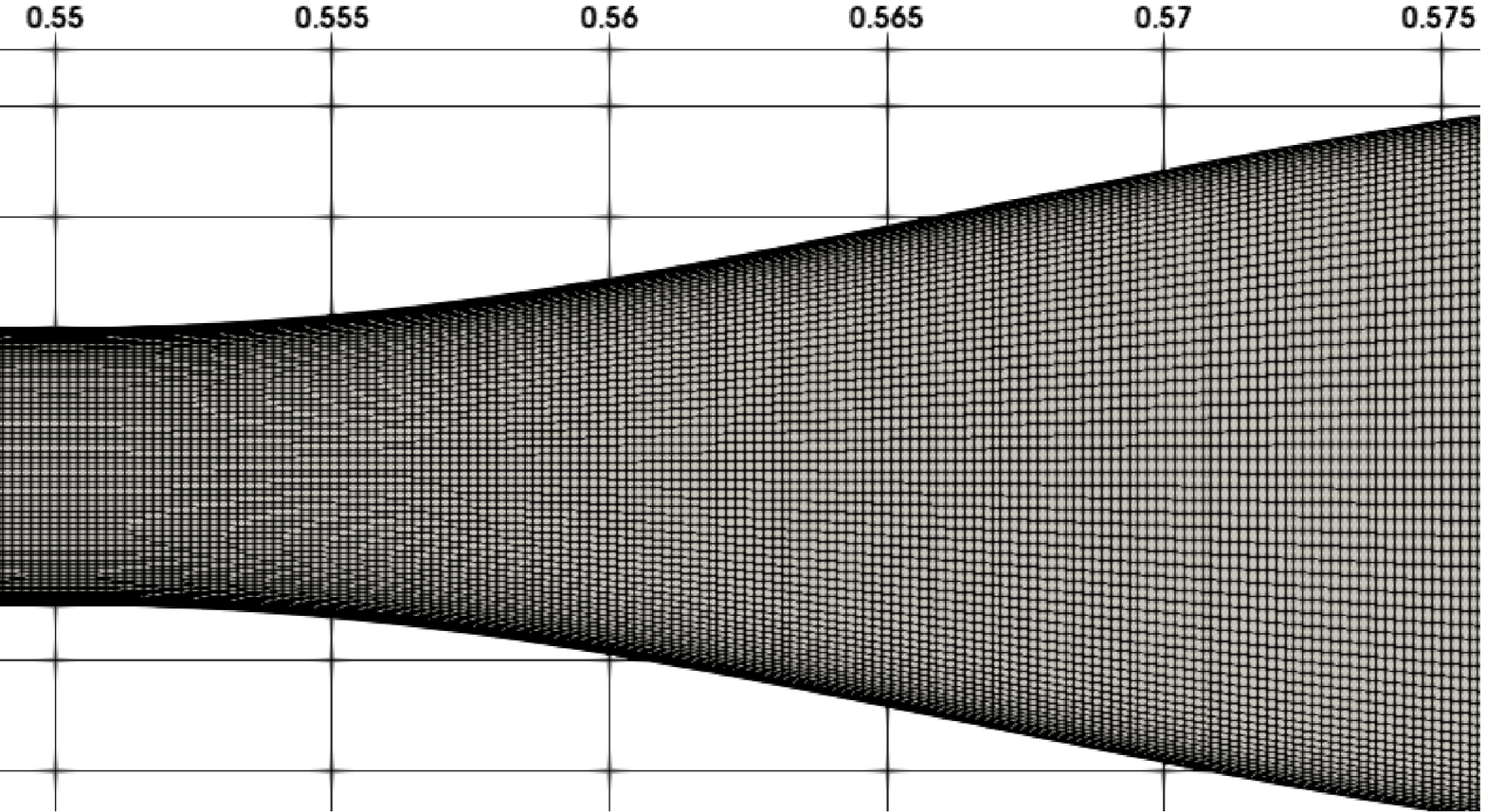}
\caption{Detail of the utilized grid for the FD study at the section $x \in [0.55-0.575]$ $m$, where $x=0$ refers to the inlet and $x=1.1$ to the outlet. (Figure in scale).  \\  \label{fig:FD_grid_detail} }
\end{minipage}
\end{figure}

The goal of the primal-adjoint simulation is the computation of the surface sensitivity (shape derivative) through Eqn. (\ref{eq:surface_sen_5}). It is thus important to investigate the accuracy of the computed sensitivity. In order to realize this, the computed shape derivative is compared against locally evaluated second order accurate finite differences \cite{papadimitriou_adj,kroger2015cad}, viz. 
\begin{equation}
    \delta_{c_i} J = \frac{\left[J(c_i + \epsilon n_i) - J(c_i - \epsilon n_i)\right]}{2\epsilon}. \label{eq:FD}
\end{equation}
Here $c_i$ represent discrete points of the control, $\epsilon$ is the magnitude of the perturbation and $n_i$ is the normal vector at $c_i$. In practise, the study is realized by deforming the boundary faces of the discretized geometry into their normal direction with a magnitude equal to $\epsilon$. The local boundary perturbations are then transported into the domain based on (\ref{eq:displ_gamma_b}).
Figure \ref{fig:FD_geometry} presents a schematic of the considered symmetric geometry. In specific, the design section ($\Gamma_D$) 
follows from $y=A\left(sin\Big(\frac{\pi}{L} x - \frac{\pi}{1000}\Big)\right)^4$, where $A$ is the height of the bump, $L$ is the length of the design surface and $L / A=20$. 
The domain is discretized with approximately $70$k control volumes on a structured grid which is refined in normal as well as tangential direction towards the wall and $\Gamma_D$. A detail of the utilized grid along a part of the design section is shown in Fig. \ref{fig:FD_grid_detail}.

\begin{figure}[ht!]
\begin{minipage}{.5\textwidth}
\centering
\includegraphics[width=1.0\textwidth]{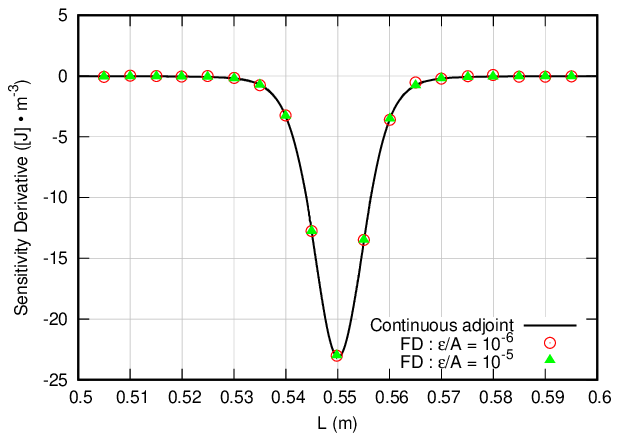}
\caption{Comparison of sensitivity predicted by Eqn. (\ref{eq:surface_sen_5}) (continuous line)
and computed from FD, Eqn. (\ref{eq:FD}), for $\epsilon / A = 10^{-6}$ (circles) and $\epsilon / A = 10^{-5}$ (triangles). 
\label{fig:FD_results}}
\end{minipage}
\hspace{5mm}
\begin{minipage}{.5\textwidth}
\centering
\includegraphics[width=1.0\textwidth]{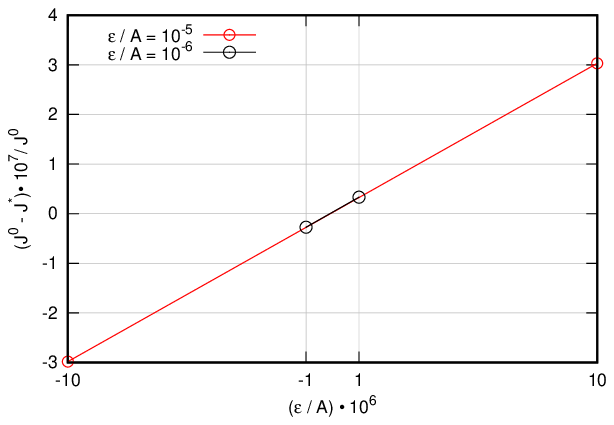}
\caption{Influence of perturbation size $\epsilon$ on FD results at $x= 0.55$ $m$. 
$J^0$ and $J^*$ represent the objective functional of the unperturbed and perturbed shapes, respectively.
 \label{fig:linear_regime_x055}}
\end{minipage}
\end{figure}

The study is conducted for a parabolic inlet velocity profile with $U_{max}=0.3$ $m/s$.
The fluid properties ($\rho, \mu$) are assigned to unity to simplify the study. The results of the FD study are shown in Fig. \ref{fig:FD_results} 
for two perturbation magnitudes, $\epsilon$, at
20 selected discrete positions. The overall agreement between the computed adjoint shape derivative and the FD results is very good. Furthermore, as shown in Fig. \ref{fig:linear_regime_x055} the computed objective functional, on the perturbed shapes, exhibits a linear behaviour w.r.t the perturbation size.

Overall, the study validates, on a preliminary basis, the accuracy of the adjoint method, as presented within this manuscript.  
\FloatBarrier

\section{Application}\label{sec:application}
Having assessed the code implementation and the reliability of the computed sensitivity, the approach is put to the test by considering a complete optimization process on a 3D geometry. The latter corresponds to a benchmark model, designed by a technical committee to include flow phenomena related to blood damage in medical devices \cite{FDA}. 

\subsection{Investigated Configuration}\label{subsec:Geometry}
The geometry under consideration shares characteristics of many blood-carrying medical devices. It includes sections where the flow is accelerating or decelerating, where re-circulation  and significant variations in shear stress occur. All of these phenomena are believed to be related to blood damage in medical devices \cite{FDA}. A merit of the adjoint method is that it stems from the primal (physical) problem and does not operate as a black box. It is therefore capable of shedding light into areas of potentially elevated blood-damage capacity. Areas on which an accumulated shape sensitivity is identified are expected to be physically relevant to the problem of hemolysis in general.

\begin{figure}[ht!]
\centering
\includegraphics[trim=0.8cm 1.5cm 0.8cm 1.5cm, width=0.8\textwidth]{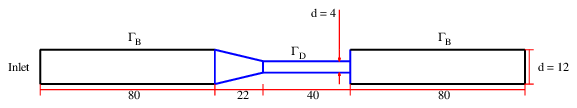}
\includegraphics[width=0.85\textwidth]{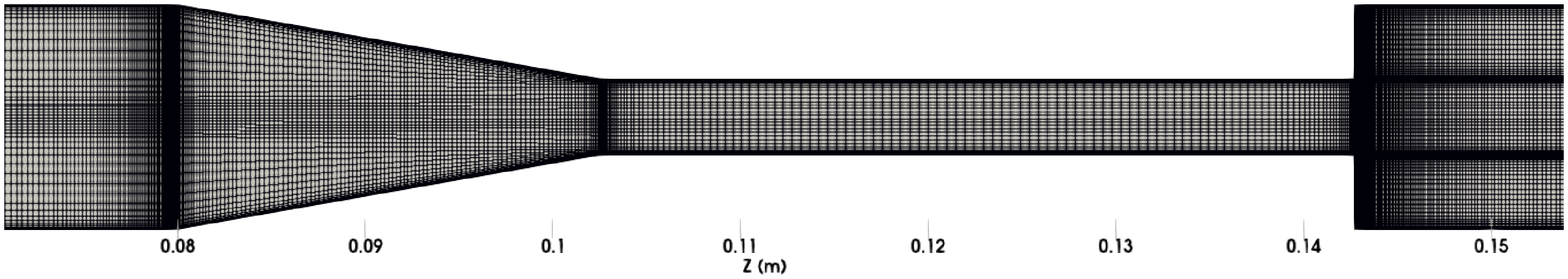}
\caption{\textbf{Top}: Sketch of the investigated geometry, adapted from \cite{FDA}. Presented in blue are the sections free for design. The remaining structure is bounded to its initial configuration. All dimensions in (mm). \\
\textbf{Bottom}: Detail of the computational grid on a longitudinal section of the geometry near the design surface. \label{fig:3D_geometry_sketch}}
\end{figure}

Geometrical details of the model are presented in Fig. \ref{fig:3D_geometry_sketch}. The wall boundary is again split into 2 parts, $\Gamma_D$ and $\Gamma_B$, as was done in the previous section. The inlet and outlet tubes are considered to be bounded, since their shape is believed to be trivial to the blood damage capacity. The remaining structure is classified as $\Gamma_D$ and presented in blue.

\begin{figure}[thp]
\begin{minipage}{.5\textwidth}
\centering
\includegraphics[width=1.0\textwidth]{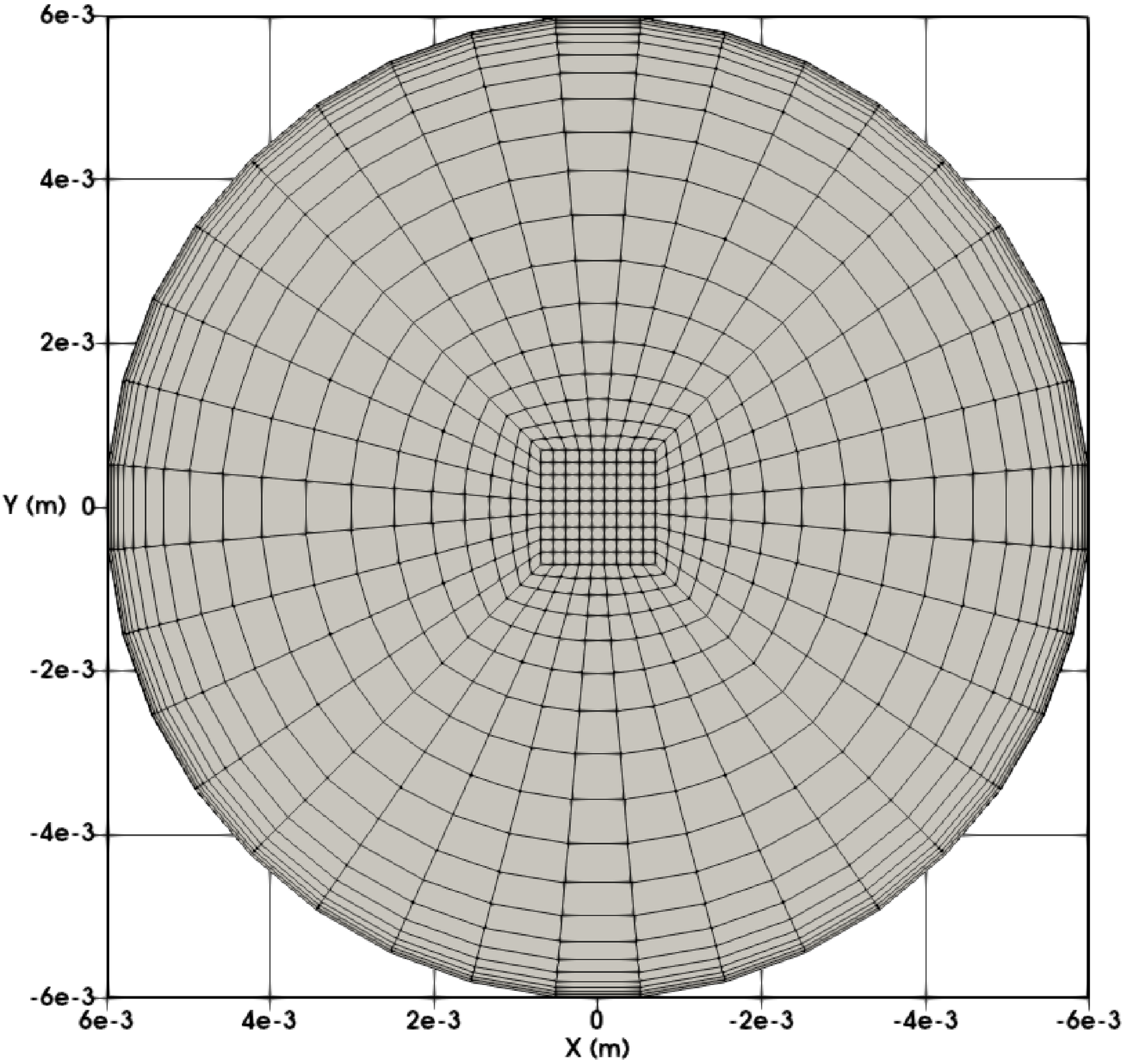}
\caption{Cross-sectional view of the grid near the inlet. \label{fig:inlet_grid}}
\end{minipage}%
\hspace{5mm}
\begin{minipage}{.5\textwidth}
\centering
\includegraphics[width=1.0\textwidth]{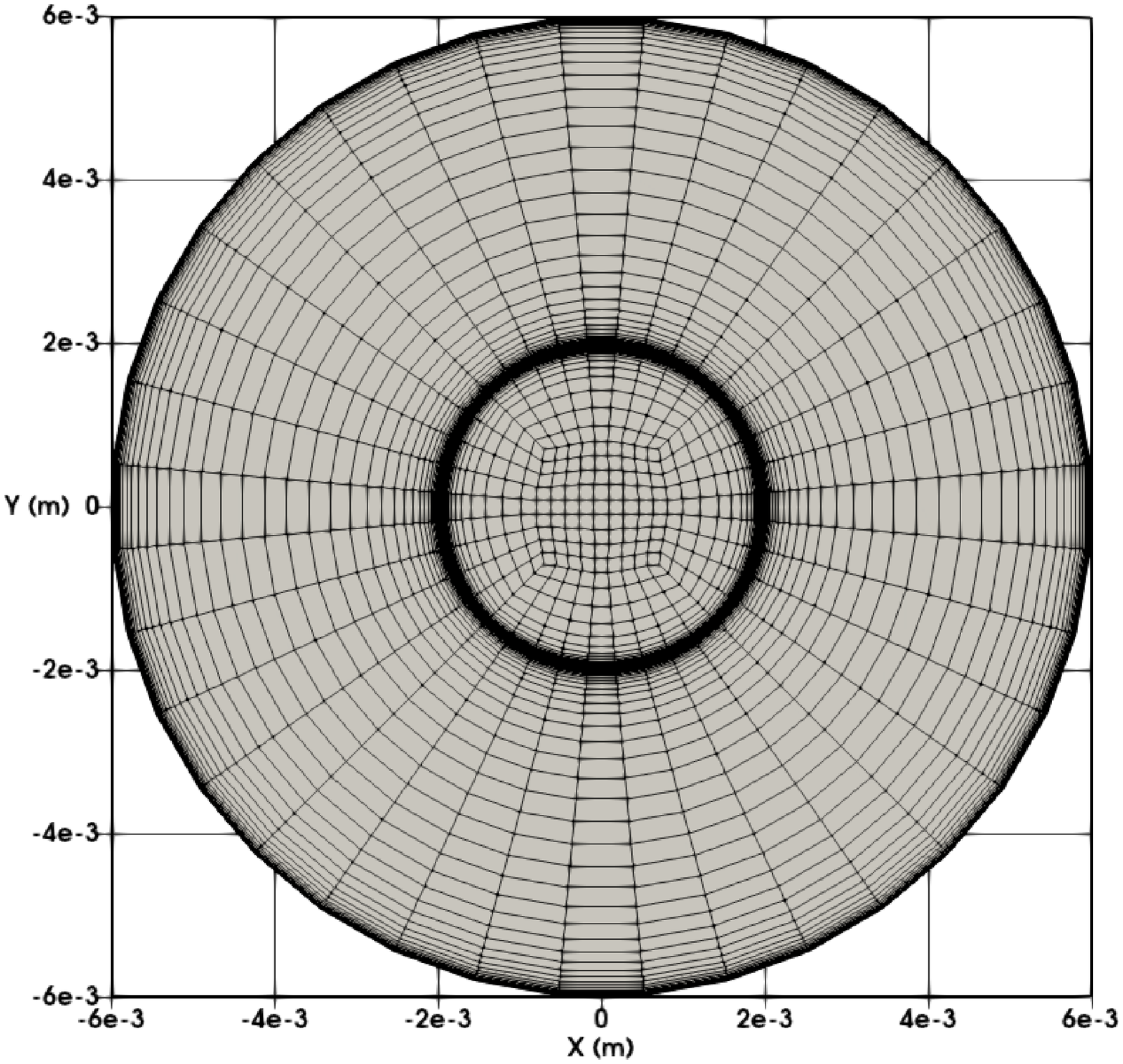}
\caption{Cross-sectional view of the grid near the outlet. \label{fig:outlet_grid} }
\end{minipage}
\end{figure}

The geometry is discretized with approximately 1 million control volumes 
on a butterfly-like structured grid. The mesh is gradually condensed near the walls, cf. Figs. \ref{fig:inlet_grid} \& \ref{fig:outlet_grid}, to adequately resolve relevant flow phenomena
and also near the $\Gamma_D$ section
(Fig. \ref{fig:3D_geometry_sketch}), to ensure an accurate computation of the sensitivity. As can be seen in Fig. \ref{fig:outlet_grid}, the grid is additionally refined in the throat region to sufficiently capture the free shear flow, occurring by the jet exiting the throat.

The fluid's density and dynamic viscosity are set to $1056$ $kg/m^3$ and $3.5$ $cP$, respectively, representing blood under physiological conditions. A fully developed laminar axial velocity profile is prescribed at the inlet of the geometry, based on (\ref{eq:pipe_fully_dev_vel}), so that the Reynolds number at the throat reads $Re_T = 500$. As regards the hemolysis-prediction model, the utilized set of parameters corresponds to GW (cf. Tab. \ref{tab:set_of_parameters}) and we employ $k=1$ for the computation of the scalar stress representation (\ref{eq:scalar_stress}). The diffusion terms in the adjoint and primal momentum equations are discretized using the second order accurate central difference scheme while the convective terms are discretized through the higher order Quadratic Upstream [Downstream] Interpolation of Convective Kinematics (QU(D)ICK) scheme.

\subsection{Shape Optimization Study}\label{subsec:results}
The optimization study  was performed for 50 design candidates. The employed smoothing parameter value (\ref{eq:gradient}) reads $\lambda = d^{T}$, where $d^{T}$ is the throat diameter. 
To avoid large deformations of the geometry and the grid, the local displacement field $d_i$ computed from Eqn. (\ref{eq:displ_gamma_b}) was scaled to a user-defined constant maximum value $\Tilde{d}_{max}$, viz.  
\begin{equation}
    \Tilde{d_i} = \frac{d_i}{max(d_i)} \Tilde{d}_{max} \label{eq:normalized_displacement}.
\end{equation}
For the study at hand $\Tilde{d}_{max}/d^{T}= 0.25 \cdot 10^{-2}$.
\FloatBarrier

\begin{figure}[ht!]
\centering
\includegraphics[width=1\textwidth]{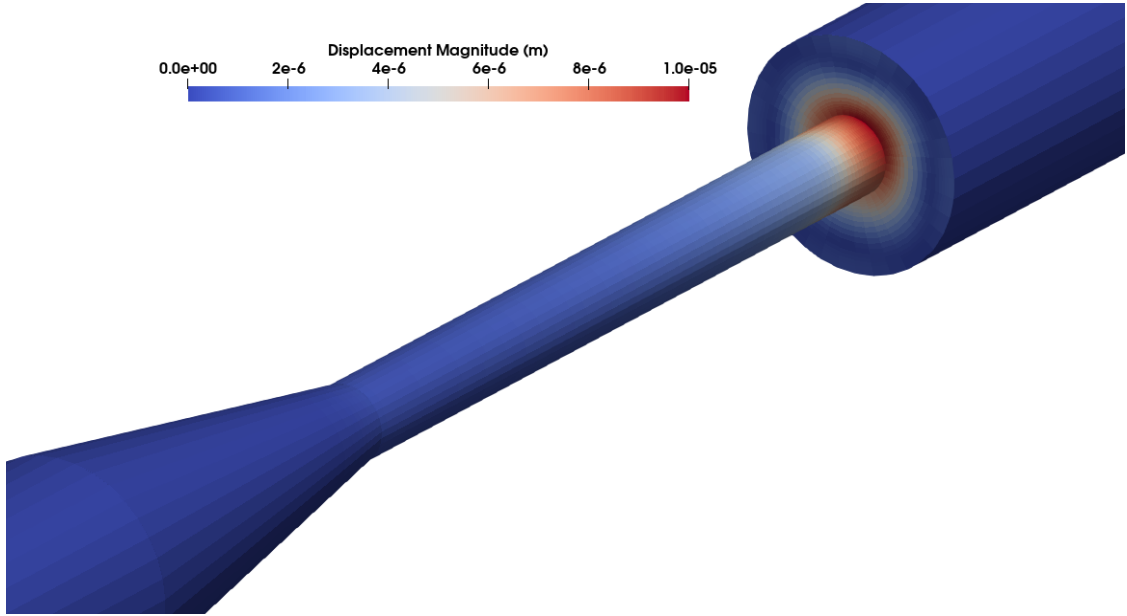}
\caption{Displacement field as computed by Eqn. (\ref{eq:normalized_displacement}) after the first primal and adjoint simulations. \label{fig:displacement_shape2}}
\end{figure}

Figure \ref{fig:displacement_shape2} shows the 
displacement magnitude on the design surfaces of the structure at hand, after the solution of the first primal and adjoint problem. As can be seen, the displacement is accumulated, almost entirely, at the sudden expansion of the geometry,
where the highest values of hemolysis occur. Furthermore, 
re-circulation is occurring with relatively significant values of upstream mass fluxes after the expansion. This is believed to be one of the causes of hemolysis. Therefore, even though there is no direct contribution from the primal or adjont hemolysis to the shape sensitivity, the necessary information from both  fields are preserved through the adjoint velocity, which is directly influenced by both $H_L$ and $\hat{h}$.

\begin{figure}[ht!]
\begin{minipage}{.5\textwidth}
\centering
\includegraphics[width=1.0\textwidth]{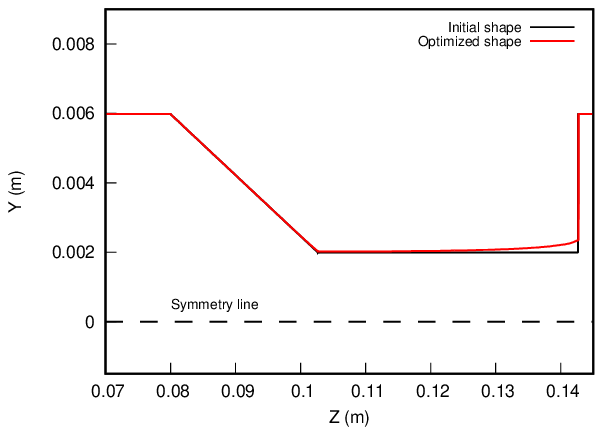}
\caption{Projection of $\Gamma_D$ on a 2D plane. Red line illustrates the outline of the final optimized shape ($c^{50}$) while black line corresponds to the initial shape. The bottom part is symmetric w.r.t the symmetry line. Axes not in scale.  \label{fig:shape_outline}}
\end{minipage}%
\hspace{5mm}
\begin{minipage}{.5\textwidth}
\centering
\includegraphics[width=1.0\textwidth]{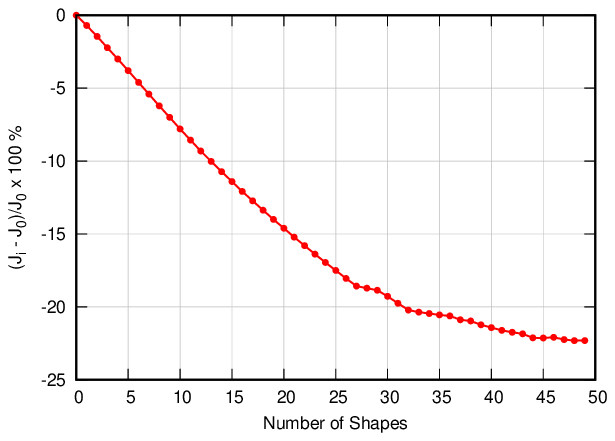}
\caption{Optimization convergence. $J_i$ denotes the objective functional as computed for each shape $c^i$ and $J_0$ denotes the same value for the initial shape, $c^0$. Points illustrate the computed quantity at each discrete shape. \label{fig:optimization_history} }
\end{minipage}
\end{figure}

\begin{figure}[ht!]
\centering
\includegraphics[scale=1.05]{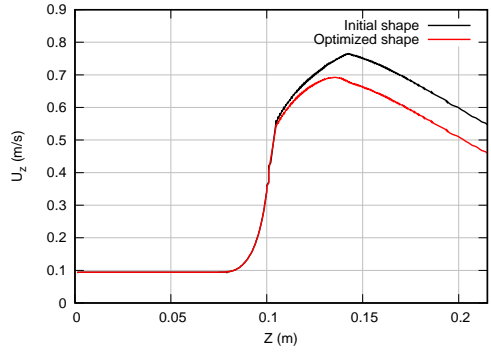}
\caption{Axial velocity along model's centerline ($r = 0$) for initial (black line) and optimized (red line) shape. Coordinate $Z = 0$ corresponds to the model's inlet. \label{fig:centerline_vel}}
\end{figure}

\begin{figure}[ht!]
\begin{minipage}{.5\textwidth}
\centering
\includegraphics[width=1.0\textwidth]{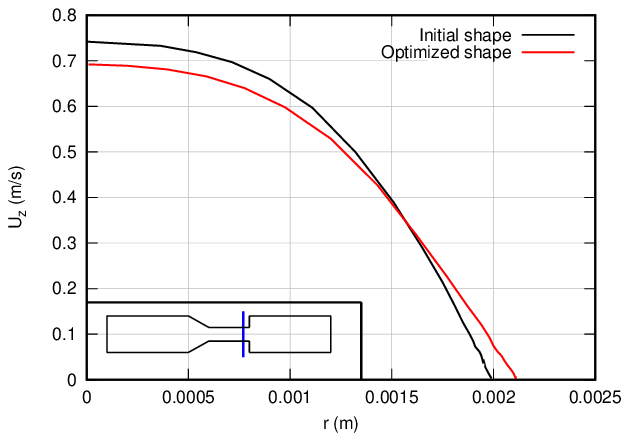}
\caption{Axial velocity cross-sectional profiles for initial (black line) and optimized (red line) shape at $Z = 0.135$ $m$. Location of the cross section is schematically shown at bottom left. \label{fig:velocity_cross_sec_1}}
\end{minipage}%
\hspace{5mm}
\begin{minipage}{.5\textwidth}
\centering
\includegraphics[width=1.0\textwidth]{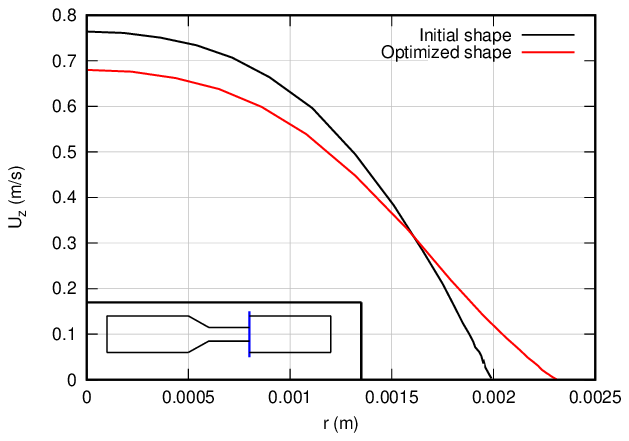}
\caption{Axial velocity cross-sectional profiles for initial (black line) and optimized (red line) shape at $Z = 0.142$ $m$, corresponding to the final nodes before the sudden expansion of the initial shape. \label{fig:velocity_cross_sec_2} }
\end{minipage}
\end{figure}

\begin{figure}[ht!]
\begin{minipage}{.5\textwidth}
\centering
\includegraphics[width=1.0\textwidth]{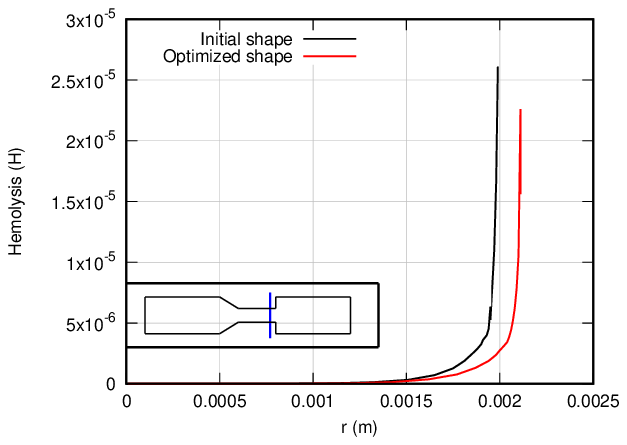}
\caption{Hemolysis cross-sectional profiles for initial (black line) and optimized (red line) shape at $Z = 0.135$ $m$. Location of the cross section is schematically shown at bottom left. \label{fig:hemolysis_cross_sec_1}}
\end{minipage}%
\hspace{5mm}
\begin{minipage}{.5\textwidth}
\centering
\includegraphics[width=1.0\textwidth]{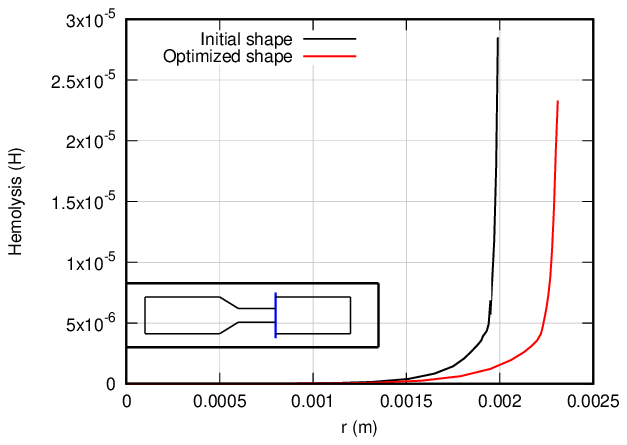}
\caption{Hemolysis cross-sectional profiles for initial (black line) and optimized (red line) shape at $Z = 0.142$ $m$, corresponding to the final nodes before the sudden expansion of the initial shape. \label{fig:hemolysis_cross_sec_2} }
\end{minipage}
\end{figure}

The optimization history is shown in Fig. \ref{fig:optimization_history}. It can be seen that the optimization starts converging after approximately 40 shapes. The final shape 
results in a relative reduction of the objective functional by 22\%. 
A comparison of the initial and the optimized shape is displayed in Fig. \ref{fig:shape_outline}. The optimization algorithm, proceeded into widening and relatively smoothing the sudden expansion of the geometry.  
This results in a smaller maximum value of the axial velocity as can be seen in Fig. \ref{fig:centerline_vel}. At the same time, nonetheless, the flow inside the bounded portions of the structure ($z \in [0:0.12]$ $m$) remains relatively unchanged as shown on the same figure. This is important in terms of biomedical applications due to their ultimate goal of realizing a sensitive task.
However, due to the smoothing of the expansion zone, the velocity profiles are changed in the perturbed section of the geometry (cf. Figs. \ref{fig:velocity_cross_sec_1} \& \ref{fig:velocity_cross_sec_2}). The velocity profile of the optimized shape is 
smoothed near the wall region resulting 
in substantially lower shear stresses.
Subsequently, due to the direct relation between shear stresses and hemolysis induction, the maximum values of hemolysis in the flow is reduced in the perturbed section of the structure, which is also the area on which the maximum values are identified in the initial shape. A direct comparison of the hemolysis profiles at two cross sections of the initial and final geometry is presented in Figs. \ref{fig:hemolysis_cross_sec_1} \& \ref{fig:hemolysis_cross_sec_2}.

\begin{figure}[ht!]
\begin{minipage}{.6\textwidth}
\centering
\includegraphics[width=1\textwidth]{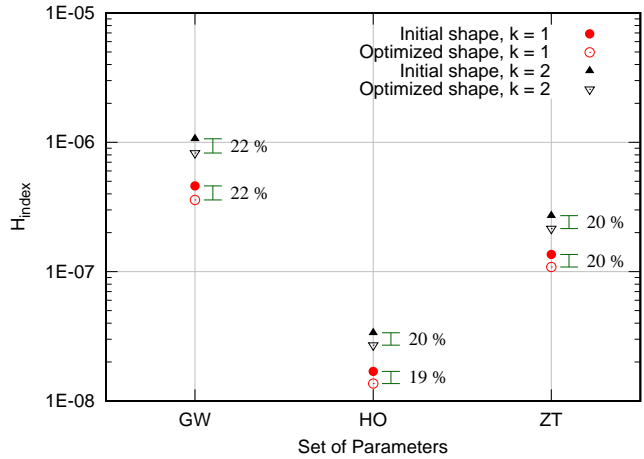}
\caption{Computed values of $H_{index}$ (Eqn. \ref{eq:objective_normalized}) for different set of parameters (cf. Tab. \ref{tab:set_of_parameters}) and different $k$ values (Eqn. (cf. \ref{eq:scalar_stress})). For each set of parameters and $k$ value the percentage decrease of the objective on the initial w.r.t to the optimized shape is shown next to points. Y-axis in logscale. \label{fig:param_study}}
\end{minipage}
\hspace{5mm}
\begin{minipage}{.4\textwidth}
\begin{tabular}[ht]{cccccr}
	\toprule
Shape                 & Param.   & $k$         & $H_{index}$  \\ \midrule
Initial  	          & GW		  & 1   &  0.460E-06  	    \\
Optimized   	      & GW 	      & 1   &   0.359E-06	    \\
Initial  	          & GW		  & 2   &  0.106E-05	    \\
Optimized   	      & GW 	      & 2   &  	0.826E-06    \\
Initial   	          & HO  	  & 1   &   0.169E-07	     \\ 
Optimized   	      & HO 	      & 1   &  0.136E-07    \\
Initial   	          & HO  	  & 2   &  0.337E-07	     \\ 
Optimized   	      & HO 	      & 2   &   0.270E-07 	    \\
Initial   	          & ZT  	  & 1   &  0.135E-06 	     \\ 
Optimized   	      & ZT 	      & 1   &  0.109E-06	    \\
Initial   	          & ZT  	  & 2   & 0.271E-06 	     \\ 
Optimized   	      & ZT 	      & 2   & 0.215E-06   	    \\\bottomrule
\end{tabular}
\\  \\ Table 3: Exact values from Fig. \ref{fig:param_study}. \\ \\ \\ \\ \\ \\ \\
\centering
\end{minipage}
\end{figure}

As described in the previous section, the optimization utilized a specific set of parameters for the primal hemolysis model. It is interesting, therefore, to examine the performance of the optimized shape for other sets of hemolysis specific parameters. As shown in Fig. \ref{fig:param_study}, for any choice of parameters mentioned in Tab \ref{tab:set_of_parameters}, the optimal shape always outperforms the initial one in terms of flow-induced hemolysis. In specific, the most significant improvement occurs for the employed set of parameters during the optimization run, namely $GW$, while the value of $k$ is relatively trivial w.r.t the improvement. This shows a robustness of the method in terms of user-defined values.

\FloatBarrier
\section{Conclusions and Outlook} \label{sec:Conclusions}
This paper discusses a continuous adjoint approach for shape optimization targeting to minimize flow-induced hemolysis in biomedical applications. A detailed derivation of the adjoint model which stems from the original power-law formulation for hemolysis prediction is reported. 

A benchmark problem is examined to verify the numerical implementation of the primal and adjoint hemolysis equations. The numerical results
compare favorably against those deduced from analytical solutions of the problem. The validity of the derived sensitivity derivative is put to the test on a two-dimensional stenosed geometry on which a FD study is realized. The continuous computed sensitivity from the adjoint method is compared against the locally evaluated results of the FD study and a fitness is found. 

The complete optimization method based on the derived adjoint equations is applied on a three-dimensional geometry, specifically designed to include flow peculiarities, related to blood damage in medical devices. An optimized shape, in 50 shape updates, able to reduce the numerically predicted flow-induced hemolysis by 22\% is found. The performance of the optimized shape, in terms of blood damage, is then tested for different hemolysis-related parameters. It is found that in all cases, the optimized geometry outperforms the initial one.

Overall, the reported method poses great potential for minimizing flow-induced hemolysis, in silico conditions, due to its computational efficiency and relative ease of numerical implementation on a standard CFD solver. While the method is derived based on a specific hemolysis prediction model on structures with rigid walls, future work will target the extension of the method to different prediction models as well as a Fluid Structure Interaction (FSI) formulation, so that wall elasticity is accounted for. 
\section*{Acknowledgments}
The current work is a part of the research training group “Simulation-Based Design Optimization
of Dynamic Systems Under Uncertainties” (SENSUS) funded by the state of Hamburg under the aegis of the Landesforschungsf{\"o}rderungs-Project LFF-GK11.
Selected computations were performed with resources provided by the North-German Supercomputing Alliance (HLRN).
This support is gratefully acknowledged by the authors.

\begin{appendices}
\section*{Appendix A}
Staying true to the notation of Sec. \ref{subsubsec:Derivation}, the derivation of the directional derivatives $I_k$, with $k=1,..,4$, is presented herein. Each sub-integral expression is repeated for the convenience of the reader. Finally, the process to arrive at the adjoint equations (Eqns. (\ref{eq:adjoint_continuity})-(\ref{eq:adjoint_hemolysis})) is also illustrated.

For $I_1$ :
\begin{equation}
       I_1(\hat{p}, u_i, \Omega) = \int_\Omega \hat{p}\Big(-\frac{\partial u_j}{\partial x_j}\Big) d\Omega \nonumber.
\end{equation}  

Since $I_1$ doesn't directly depend on $p$ and $H_L$ it follows

\begin{equation}
    \delta_p I_1 \cdot \delta p = \delta_{H_L} I_1 \cdot \delta H_L = 0 \label{eq:I1_var_p_H},
\end{equation}

while for the variation in the direction of the velocity components, by changing the index $j$ to $i$ and performing integration by parts, we get

\begin{align}
    \delta_{u_i} I_1 \cdot \delta u_i &= -\int_\Omega \hat{p}\frac{\partial \delta u_i}{\partial x_i}d\Omega \nonumber\\
                                      &=-\int_\Gamma (\delta u_i n_i) \hat{p} d\Gamma +\int_\Omega (\delta u_i) \frac{\partial \hat{p}}{\partial x_i} d\Omega \label{eq:I1_var_u}.
\end{align}

For $I_2$ :

\begin{equation}
       I_2(\hat{u}_i, u_i, \Omega) = \int_\Omega \hat{u}_i \Big(\rho u_j\frac{\partial u_i}{\partial x_j}\Big) d\Omega \nonumber.
\end{equation}  

Since $I_2$ doesn't directly depend on $p$ and $H_L$ it follows

\begin{equation}
    \delta_p I_2 \cdot \delta p = \delta_{H_L} I_2 \cdot \delta H_L = 0 \label{eq:I2_var_p_H}
\end{equation}

and for its variation on the direction of the velocity components

\begin{align}
    \delta_{u_i} I_2 \cdot \delta u_i  & =  \int_\Omega \hat{u}_i\rho\Big( u_j\frac{\partial \delta u_i}{\partial x_j} + \delta u_j \frac{\partial u_i}{\partial x_j} + \delta u_j \frac{\partial \delta u_i }{\partial x_j} \Big) d\Omega \nonumber \\
    & =  \int_\Omega \hat{u}_i\rho u_j\frac{\partial \delta u_i}{\partial x_j} d\Omega + \int_\Omega \hat{u}_j\rho \delta u_i \frac{\partial u_j}{\partial x_i} d\Omega+ \overbrace{HoT}^{neglect} \nonumber \\
     & =  \int_\Gamma \hat{u}_i\rho (u_j n_j) \delta u_i d\Gamma - \int_\Omega (\delta u_i) \rho u_j \frac{\partial \hat{u}_i}{\partial x_j} d\Omega \nonumber \\
     &+\int_\Omega \hat{u}_j \rho (\delta u_i) \frac{\partial u_j}{\partial x_i} d\Omega \nonumber \\
     &= \int_\Gamma \delta u_i \hat{u}_i \rho (u_j n_j) d\Gamma - \int_\Omega (\delta u_i) \left(u_j \rho \frac{\partial \hat{u}_i}{\partial x_j} - \hat{u}_j \rho \frac{\partial u_j}{\partial x_i}\right) d\Omega.
     \label{eq:I2_var_u}
\end{align}

The volume integral on the 5th line of Eqn. (\ref{eq:I2_var_u}) could also be integrated by parts. This would lead to an exchange of the adjoint and primal velocities on the volume integral and an additional surface integral. However, neglecting this operation could stabilize the numerical solution of the adjoint system \cite{Soto2004OnTC}. In this study, the derivation as presented in Eqn. (\ref{eq:I2_var_u}) is solely considered.

For $I_3$ :

\begin{equation}
I_3(\hat{u}_i, u_i, \Omega) =\int_\Omega \hat{u}_i\Big(-\frac{\partial}{\partial x_j}(2\mu S_{ij})\Big) d\Omega \nonumber.
\end{equation}

As before

\begin{equation}
    \delta_p I_3 \cdot \delta p = \delta_{H_L} I_3 \cdot \delta H_L = 0 \label{eq:I3_var_p_H}
\end{equation}

\begin{align}
    \delta_{u_i} I_3 \cdot \delta u_i  & =  - \int_\Omega \hat{u}_i\frac{\partial}{\partial x_j}\left[ \mu \Big(\frac{\partial \delta u_i}{\partial x_j} + \frac{\partial \delta u_j}{\partial x_i}\Big)\right] d\Omega \nonumber \\
    & =\underbrace{ - \int_\Gamma \hat{u}_i \mu \Big(\frac{\partial \delta u_i}{\partial x_j} + \frac{\partial \delta u_j}{\partial x_i}\Big) n_j d\Gamma}_{\Gamma_1} + \underbrace{\int_\Omega \frac{\partial\hat{u}_i}{\partial x_j} \mu \Big(\frac{\partial \delta u_i}{\partial x_j} + \frac{\partial \delta u_j}{\partial x_i}\Big)d\Omega}_{I_3^*}
     \label{eq:I3_var_u_1},
\end{align}

where $I_3^*$ can be expanded further using a second integration by parts as

\begin{align}
    I_3^* &= \int_\Omega \frac{\partial \hat{u}_i}{\partial x_j} \mu \frac{\partial \delta u_i}{\partial x_j} d \Omega + \int_\Omega \frac{\partial \hat{u}_i}{\partial x_j} \mu \frac{\partial \delta u_j}{\partial x_i} d \Omega \nonumber \\
    &= \int_\Gamma (\delta u_i) \mu \frac{\partial \hat{u}_i}{\partial x_j} n_j d\Gamma - \int_\Omega \delta u_i \frac{\partial}{\partial x_j}\left[\mu \frac{\partial \hat{u}_i}{\partial x_j}\right] d\Omega \nonumber \\
    &+ \int_\Gamma (\delta u_j) \mu \frac{\partial \hat{u}_i}{\partial x_j} n_i d\Gamma - \int_\Omega \delta u_j \frac{\partial}{\partial x_i}\left[\mu \frac{\partial \hat{u}_i}{\partial x_j}\right] d\Omega \nonumber \\
    &= \underbrace{\int_\Gamma (\delta u_i) \mu \left[\frac{\partial \hat{u}_i}{\partial x_j}+\frac{\partial \hat{u}_j}{\partial x_i}\right]n_j d \Gamma}_{\Gamma_2} - \int_\Omega (\delta u_i) \frac{\partial}{\partial x_j}\left[\mu \Big(\frac{\partial \hat{u}_i}{\partial x_j}+\frac{\partial \hat{u}_j}{\partial x_i}\Big)\right] d \Omega.
    \label{eq:I3_var_u_2}
\end{align}

$\Gamma_1 + \Gamma_2$ are subject to even further expansion

\begin{align}
    \Gamma_1 + \Gamma_2 &= \int_\Gamma \left[(\delta u_i) \mu \Big[\frac{\partial \hat{u}_i}{\partial x_j}+\frac{\partial \hat{u}_j}{\partial x_i}\Big]n_j - \hat{u}_i\mu\Big[\frac{\partial \delta u_i}{\partial x_j} + \frac{\partial \delta u_j}{\partial x_i}\Big]n_j\right] d\Gamma \nonumber \\
    &= \int_\Gamma \mu n_j \left( \frac{\partial \hat{u}_j}{\partial x_i} \delta u_i - \frac{\partial \delta u_j}{\partial x_i}\hat{u}_i\right) d\Gamma + \int_\Gamma \mu n_j \left( \frac{\partial \hat{u}_i}{\partial x_j} \delta u_i - \frac{\partial \delta u_i}{\partial x_j}\hat{u}_i\right) d\Gamma \label{eq:I3_var_u_3}.
\end{align}

Expanding the first integral on Eqn. (\ref{eq:I3_var_u_3}) leads to

\begin{align}
    &\int_\Gamma \mu n_j \left( \frac{\partial \hat{u}_j}{\partial x_i} \delta u_i - \frac{\partial \delta u_j}{\partial x_i}\hat{u}_i\right) d\Gamma = \int_\Gamma \mu \frac{\partial \hat{u}_n}{\partial x_i} \delta u_i d \Gamma - \int_\Gamma \mu \frac{\partial \delta u_n}{\partial x_i}\hat{u}_i d\Gamma \nonumber \\
    &=\left[\mu \hat{u}_n\underbrace{\delta u_i n_i}_{\delta u_n} \right]_{\partial \Gamma} - \left[\mu \underbrace{\hat{u}_i n_i}_{\hat{u}_n} \delta u_n \right]_{\partial \Gamma} - \int_\Gamma \hat{u}_n \mu \frac{\partial \delta u_i}{\delta x_i} d\Gamma + \int_\Gamma \delta u_n \mu \frac{\partial \hat{u}_i}{\partial x_i} d\Gamma \nonumber \\
    & = 0 \label{eq:I3_var_u_4}.
\end{align}

The last two integrals of Eqn. (\ref{eq:I3_var_u_4}) vanish because the primal velocity field is asymptotically divergence-free in the boundary so its variation must be divergence-free too and because we showed that the adjoint velocity must also be divergence-free. Finally, simplifying Eqn. (\ref{eq:I3_var_u_3}) through Eqn. (\ref{eq:I3_var_u_4}) and inserting the remaining terms in Eqn. (\ref{eq:I3_var_u_1}) we get

\begin{align}
    \delta_{u_i} I_3 \cdot \delta u_i  & = \int_\Gamma (\delta u_i) \mu n_j \frac{\partial \hat{u}_i}{\partial x_j} d\Gamma - \int_\Gamma \hat{u}_i \mu n_j \frac{\partial \delta u_i}{\partial x_j} d\Gamma  - \int_\Omega (\delta u_i) \frac{\partial}{\partial x_j} \left(2 \mu \hat{S}_{ij}\right) d \Omega
\label{eq:I3_var_u_final}.
\end{align}

For $I_4$ :

\begin{equation}
I_4(\hat{u}_i, p, \Omega)  = \int_\Omega \hat{u}_i\Big(\frac{\partial p}{\partial x_i}\Big) d\Omega \nonumber.
\end{equation}

It follows that

\begin{equation}
    \delta_{u_i} I_4 \cdot \delta u_i = \delta_{H_L} I_4 \cdot \delta H_L = 0 \label{eq:I4_var_u_H},
\end{equation}

while for the variation in the direction of pressure we can write
\begin{align}
    \delta_{p} I_4 \cdot \delta p  & = \int_{\Omega} \hat{u}_i\frac{\partial \delta p}{\partial x_i} \delta \Omega \nonumber \\
    &=  \int_\Gamma (\delta p) \hat{u}_i n_i d\Gamma - \int_{\Omega} (\delta p) \frac{\partial \hat{u}_i}{\partial x_i} d\Omega
\label{eq:I4_var_p}.
\end{align}

Having expressed all sub-integrals in the form of surface and volume integrals we can superpose all contributions (considering also those from the main body of the paper) into the directional derivatives of the Lagrange functional so that

\begin{equation}
    \delta_p \pazocal{L} \cdot \delta p = \int_\Gamma (\delta p) \hat{u}_i n_i d\Gamma - \int_{\Omega} (\delta p) \frac{\partial \hat{u}_i}{\partial x_i} d\Omega \label{eq:L_var_p}
\end{equation}
    
\begin{align}
   \delta_{u_i}L \cdot \delta u_i  &= \int_\Gamma (\delta u_i) \left[-\hat{p} n_i + \rho \hat{u}_i u_j n_j + \mu \frac{\partial \hat{u}_i}{\partial x_j} n_j + H_L \hat{h} n_i - \hat{h}B_{ij}n_j + \frac{\partial j_\Gamma}{\partial u_i}\right] d\Gamma \nonumber \\
    & + \int_\Omega (\delta u_i) \left[\hat{u}_j \rho \frac{\partial u_j}{\partial x_i} - u_j \rho \frac{\partial \hat{u}_i}{\partial x_j}-\frac{\partial}{\partial x_j}\Big(2 \mu \hat{S}_{ij} - \hat{p}\delta_{ij}\Big) + \frac{\partial (\hat{h}B_{ij})}{\partial x_j} - H_L\frac{\partial \hat{h}}{\partial x_i} \right] d\Omega \nonumber \\
    & - \int_\Gamma \mu \hat{u}_i \frac{\partial \delta u_i}{\partial x_j}n_j d\Gamma \label{eq:L_var_u}
\end{align}

\begin{align}
    \delta_h L \cdot \delta h = \int_\Gamma (\delta h) \Big(\hat{h}(u_j n_j) + \frac{\partial j_\Gamma}{\partial H_L} \Big) d\Gamma + \int_\Omega (\delta h) \Big( u_j \frac{\partial \hat{h}}{\partial x_j} - C^{\frac{1}{\beta}} \overline{\tau}^{\frac{a}{b}}\hat{h} \Big) d\Omega \label{eq:L_var_H}.
\end{align}

Finally, by demanding that the integrals of the form of Eqn. (\ref{eq:lagrange_variation_y}) disappear for every test function, $\delta y$, one arrives at the adjoint equations (Eqns. (\ref{eq:adjoint_continuity}) - (\ref{eq:adjoint_hemolysis})) as well as their BCs (Eqns. (\ref{eq:adjoint_hemolysis_BC_2})). It is worth noting that the second surface integral on the RHS of Eqn. (\ref{eq:L_var_u}) does not conform to the integrals of Eqn. (\ref{eq:lagrange_variation_y}). This term might contribute to the sensitivity derivative depending on the adjoint BCs.

\section*{Appendix B}
\subsubsection*{Additional information for the computation of surface sensitivity}
The mathematical manipulations that are followed for the derivation of Eqn. (\ref{eq:surface_sen_5}) from Eqn. (\ref{eq:surface_sen_4}) are presented herein. The latter equation is rewritten here for the convenience of the reader

\begin{equation*}
\delta_c \pazocal{L} \cdot \delta c = - \int_\Omega \hat{y} \Big(\delta_y R \cdot \delta y \Big) d\Omega.
\end{equation*}

The RHS of the equation can also be written as

\begin{equation}
    RHS = - \left[\delta_y I_1 \cdot \delta y + \delta_y I_2 \cdot \delta y + ... + \delta_y I_6 \cdot \delta y\right] \label{eq:appendix_B_1}.
\end{equation}

However, the quantities $\delta_y I_k \cdot \delta y$ are already estimated for the derivation of the adjoint equations. In specific, in App. A, it is shown that the quantity inside brackets in Eqn. (\ref{eq:appendix_B_1}) is equal to $\Big(\delta_p \pazocal{L} \cdot \delta p + \delta_{u_i} \pazocal{L} \cdot \delta u_i + \delta_{H_L} \pazocal{L} \cdot \delta H_L - \delta_y J \cdot \delta y\Big)$, where the expressions of the variations are presented in Eqns. (\ref{eq:L_var_p}) -(\ref{eq:L_var_H}). Considering that the surface sensitivity is defined only in $\Gamma_{D} \subset \Gamma_{W}$, all the contributions from volume integrals vanish. Therefore, what remains is

\begin{align}
    RHS &= - \Bigg[\int_{\Gamma_D} (\delta p) \hat{u}_i n_i d\Gamma - \int_{\Gamma_D} \mu \hat{u}_i \frac{\partial \delta u_i}{\partial x_j}n_j d\Gamma + \int_{\Gamma_D} (\delta h) \Big(\hat{h}(u_j n_j) \Big) d\Gamma \nonumber \\
    &+ \int_{\Gamma_D} (\delta u_i) \Big(-\hat{p} n_i + \rho \hat{u}_i u_j n_j + \mu \frac{\partial \hat{u}_i}{\partial x_j} n_j + H_L \hat{h} n_i - \hat{h}B_{ij}n_j \Big) d\Gamma \Bigg] \label{eq:appendix_B_2}.
\end{align}

By taking into account the BCs of the problem at hand (Tab. \ref{tab:Boundary_Conditions} in Sec. \ref{subsec:BC}), one can see that most terms of Eqn. (\ref{eq:appendix_B_2}) vanish and RHS is simplified to

\begin{equation}
    RHS = - \int_{\Gamma_D} (\delta u_i) \left[\mu \frac{\partial \hat{u}_i}{\partial x_j} n_j - \hat{p}n_i\right] d\Gamma \label{eq:appendix_B_3}.
\end{equation}

At this point, it is important to mention that the deformations $\delta c$ correspond to small variations in the normal direction of the shape. Based on that, we take a well-know approximation \cite{Heners,Soto2004OnTC}, that employs a Taylor expansion on the velocity field, to write

\begin{equation}
     \delta u_i = \frac{\partial u_i}{\partial n} \delta c + \underbrace{O \left((\delta c)^2\right)}_{HoT} \approx \frac{\partial u_i}{\partial n} \delta c \label{eq:appendix_B_4}.
\end{equation}

Substituting Eqn. (\ref{eq:appendix_B_4}) to (\ref{eq:appendix_B_3}) we get

\begin{align}
     \delta_{c}L \cdot \delta c &= - \int_{\Gamma_W} \frac{\partial u_i}{\partial n}\delta c \left[\mu n_j \frac{\partial \hat{u}_i}{\partial x_j} - \hat{p}n_i\right] d \Gamma  \nonumber \\
     &= - \int_{\Gamma_W} \delta c \left[\mu \frac{\partial u_i}{\partial n}\frac{\partial \hat{u}_i}{\partial n}\right]d\Gamma \label{eq:appendix_B_5},
\end{align}

where the last term of Eqn. (\ref{eq:appendix_B_5}) vanishes because the primal continuity equation asymptotically holds on the boundaries of the domain also. Finally, if we consider the velocity components as the summation of their normal and tangential parts, one arrives at the expression of Eqn. (\ref{eq:surface_sen_5}).

\end{appendices}

\bibliography{Biblio.bib}
\bibliographystyle{abbrv}
\end{document}